\documentclass[12pt]{iopart}

\usepackage[square,numbers,sort&compress]{natbib}

\usepackage{hyperref}
\hypersetup{
    colorlinks,
    citecolor=blue,
    filecolor=black,
    linkcolor=blue,
    urlcolor=blue
}

\usepackage{amsfonts}
\usepackage[backgroundcolor=white,textsize=tiny]{todonotes}

\begin{document}

\title{Perspective on new implementations of  atomtronic circuits}

\author{Juan Polo$^{1*}$, Wayne J. Chetcuti$^1$, Enrico Domanti$^{1,2,3}$, Philip Kitson$^{1,2,3}$, Andreas Osterloh$^1$, Francesco Perciavalle$^{1,4}$, Vijay Pal Singh$^1$, Luigi Amico$^{1,2,3,5}$}
\address{$^1$Quantum Research Center, Technology Innovation Institute, P.O. Box 9639 Abu Dhabi, UAE}
\address{$^2$Dipartimento di Fisica e Astronomia, Via S. Sofia 64, 95127 Catania, Italy}
\address{$^3$INFN-Sezione di Catania, Via S. Sofia 64, 95127 Catania, Italy}
\address{$^4$Dipartimento di Fisica dell’Universita` di Pisa and INFN, Largo Pontecorvo 3, I-56127 Pisa, Italy}
\address{$^5$Centre for Quantum Technologies, National University of Singapore 117543, Singapore}
\ead{$^*$juan.polo@tii.ae}
\vspace{10pt}

\begin{abstract}

In this article, we provide perspectives for atomtronics circuits on quantum technology platforms beyond simple bosonic or fermionic cold atom matter-wave currents.
Specifically, we consider (i) matter-wave schemes with multi-component quantum fluids; (ii) networks of Rydberg atoms that provide a radically new concept of atomtronics circuits in which the flow, rather than in terms of matter, occurs through excitations;
(iii) hybrid matterwave circuits - cavities systems that can be used   to study atomtronic  circuits beyond the  standard solutions and provide new schemes for integrated matter-wave networks. We also sketch how driving these systems can open new pathways for atomtronics.    
\end{abstract}

\vspace{2pc}

\submitto{Quantum Science and Technology}

\section{Introduction}

Advancements in micro-optics technology gifted us with various techniques to craft and sculpt magnetic fields and light in 
a plethora of architectures~\cite{gauthier2016direct,rubinsztein2016roadmap,henderson2009experimental}. In such magneto-optical traps, ultracold atoms can be coherently trapped and manipulated with unprecedented control of their physical conditions~\cite{bloch2005ultracold,cornell2002nobel,ketterle2002nobel,dalfovo1999theory}. These tailored and flexible microscopic magneto-optical potentials are the bedrock of atomtronics, the quantum technology of guided cold atoms~\cite{seaman2007atomtronics,amico2017focus,amico2021roadmap,amico2022colloquium}. 
Capitalizing on the versatility and customizability of these landscapes in conjunction with the inherent characteristics of cold atoms, atomtronics seeks to construct circuits of neutral matter-waves bringing to fruition a long-standing goal of the field~\cite{dekker2000guiding,dumke2002interferometer,leanhardt2002propagation,muller2000guiding,schneble2003integrated,denschlag1999guiding,denschlag1999neutral,schmiedmayer1995wire,schmiedmayer1996classical,schmiedmayer1996particle,schmiedmayer1995guiding}.

An important motivation of the field has been to emulate classical electronic devices like diodes and transistors~\cite{pepino2021entropy,pepino2009atomtronic,gou2020tunable,stickney2007transistorlike,caliga2016principles,mclain2018quantum,wilsmann2018control,zozulya2013principles,caliga2017experimental,caliga2012matterwave,caliga2016transport} or Josephson junction-based circuital elements such as SQUIDs~\cite{amico2005quantum,wright2013driving,aghamalyan2015coherent,haug2018readout,ryu2020quantum,pezze2023stabilizing,wang2015resonant,matthew2015self,yakimenko2015vortices,mehdi2021superflow,mathey2016realizing,rastelli2013quantum,schenke2011nonadiabatic,amico2014superfluid,solenov2010macroscopic,hallwood2006macroscopic,nunnenkamp2008generation,safaei2018two}. Already in the early vision~\cite{amico2005quantum}, atomtronics sets a new logic for cold atoms quantum simulators~\cite{bloch2008many,buluta2009quantum,cirac2012goals,dowling2003quantum,lamata2014epj,lewenstein2012ultracold}: mirroring the $I-V$ characteristics in solid-state physics atomtronics proposes to harness neutral matter-wave currents to diagnose fundamental questions in mesoscopic quantum matter.

Various studies have been carried out with the general mindset of interpolating basic quantum science with a more device-oriented mission~\cite{amico2021roadmap,amico2022colloquium}. Specifically, persistent currents in ring-shaped bosonic and more recently fermionic systems have been thoroughly analyzed theoretically and explored experimentally both for studying fundamental features of quantum coherent matter~\cite{moulder2012quantized,pandey2019hypersonic,guo2020supersonic,pace2022imprinting,cai2022persistent,kumar2017temperature,jendrzejewski2014resistive,eckel2014hysteresis,cominotti2014optimal,hekking1997quantum,matveev2002persistent,haug2019machine,aghamalyan2013effective,haug2018mesoscopic,pecci2021probing,pecci2023single,polo2016transport,richaud2017quantum,polo2018damping,victorin2019nonclassical,lesanovsky2007time,nicolau2020orbital,oliinyk2019tunneling,danshita2012quantum,kunimi2017thermally,eller2020producing,murray2013probing,corman2014quench,roscilde2016from,waintal2008persistent,yu1992persistent,patu2022temperature,arabahmadi2021universal,dubessy2012critical,wright2013threshold,mathey2014decay,wang2015resonant,abad2016persistent,burchianti2018connecting,polo2019oscillations,naldesi2019rise,xhani2020critical,naldesi2022enhancing,perezobiol2022coherent,kumar2017temperature,singh2020josephson,wlazlowski2023dissipation, klejdja2023decay}
and for their possible exploitation as rotation sensors ~\cite{wright2013driving,ramanathan2011superflow,eckel2014interferometric,ryu2013experimental,aghamalyan2015coherent,polo2022quantum,piazza2009vortex}. 
Working principles of guided interferometry have been 
drawn~\cite{qi2017magnetically,wu2007demonstration,burke2009scalable,burke2008confinement,deissler2008measurement,arimondo2009atom,bongs2019taking,geiger2020high,geiger2011detecting,muller2008atom,stockton2011absolute,sugarbaker2014atom,kovachy2015matter,muntinga2013interferometry,helm2015sagnac,helm2012bright,mcdonald2014bright,polo2013soliton,marchukov2019splitting,krzyzanowska2022matter,kim2022one,weiss2009creation,streltsov2009scattering,haine2018quantum,naldesi2022massive,riedel2010atom,jo2007phase,schumm2005matter,gunther2007atom,bohi2009coherent,akatsuka2017optically,ryu2015integrated,moan2020quantum,navez2016matter,stevenson2015sagnac,halkyard2010rotational,japha2007using,pelegri2018quantum,marti2015collective,moukouri2021multipass}. Relevant aspects of mesoscopic quantum transport have been thoroughly 
explored~\cite{gauthier2019quantitative,lebrat2018band,lebrat2019quantized,krinner2014observation,krinner2017two,sommer2011universal,brantut2012conduction,brantut2013thermoelectric,husmann2015connecting,haug2019aharonov,haug2019andreev,burchianti2018connecting,delpace2021tunneling,stadler2012observing,valtolina2015josephson,barontini2013controlling,labouvie2015negatice,pezze2018quantum,haroche2006exploring,leBlanc2011dynamics,spagnolli2017crossing,albiez2005direct,levy2007the,pigneur2018relaxation,gati2006noise,trenkwalder2016quantum,gaunt2012robust,mcgloin2003applications,daley2008andreev,watabe2008reflection,zapata2009andreev,tokuno2008dynamics,damanet2019controlling,corman2019quantized,mullers2018coherent,filippone2016violation,husmann2018breakdown,grenier2014peltier,sekera2016thermoelectricity,pershoguba2019thermopower,celi2014synthetic,uchino2017anomalous,Kanasz2016anomalous,liu2017anomalous,yao2018controlled,gutman2012cold,simpson2014one,salerno2019quantized,you2019atomtronics,price2017synthetic,hausler2017scanning,white2019observation,bruderer2012mesoscopic,chien2014landauer,gallego2014nonequilibrium,ivanov2013bosonic,kolovsky2017microscopic,nietner2014transport,kolovsky2018landauer,papoular2016quantized,cuevas2017molecular,krinner2013superfluidity,krinner2015observation,krinner2015observationa,pyykkonen2021flat,kwon2020strongly,zaccanti2019critical,uchino2020bosonic,uchino2020role,luick2020ideal,tononi2020dephasing,binanti2021dissipation,schweigler2021decay,langen2015ultracold,betz2011two,gring2012relaxation,hofferberth2007nonequilibrium,hofferberth2008probing,giovanazzi2008effective}. 

In this perspective article, we want to highlight a core aspect of the field: relying on the progress in cold atoms technology, the quantum fluid coursing through atomtronic networks can be of different nature ranging from weakly to strongly interacting particles, from  short-to-long range,  and with different statistical properties.
The resulting systems enable blueprints for various devices and simulators with distinct specifications, performances, and functionalities. 
Specifically, we focus on atomtronic circuits of multi-component gases, Rydberg atoms and hybrid devices, providing platforms beyond the standard bosonic or fermionic realizations. We will discuss how such circuits can provide new applications in quantum technology, for instance as interferometers, quantum simulators for high-energy physics and driven atomtronic circuits.

\section{Multi-component gases}

With the recent experimental developments, the field has progressed to a point where atomtronic networks of ultracold fermions can now be explored~\cite{cai2022persistent,pace2022imprinting}. Relying on the remarkable experimental achievements in the coherent control of $N$-component fermions/bosons~\cite{fukuhara2007degenerate,fukuhara2007bose,fukuhara2009all,cazalilla2009ultracold}, further progress can be envisaged. Indeed, the different internal states in the system add an extra level of complexity compared to a single-component system~\cite{ho1998spinor,ohmi1998bose,yip1999zero,stamper2013spinor,cazalilla2014ultracold,mistakidis2023few} that can set 
atomtronic circuits with unique functionalities and specifications. An example is that of SU($N$) fermions, as provided by alkaline earth-like gases~\cite{gorshkov2010two,cazalilla2014ultracold,capponi2016phases}, which are relevant both for quantum simulation~\cite{scazza2014observation,hofrichter2016direct,kolkowitz2016spin,taie2022observation,takahashi2022quantum} and for high-precision measurements~\cite{ludlow2015optical,marti2018imaging}.

A natural direction for multi-component atomtronics is that of quantum transport~\cite{amico2021roadmap,amico2022colloquium}. Specifically, one could focus on fermionic transport, which, whilst extensively investigated in solid-state physics~\cite{imry2002introduction}, could be revisited in  new ways, owing to the specific features of the  cold atoms technology~\cite{bloch2008many}.
With a similar logic as that applied for two-component fermions~\cite{levy2007the,luick2020ideal,kwon2020strongly,delpace2021tunneling,giorgini2008theory,pecci2021probing}, relevant problems like Josephson currents and the BEC-BCS crossover~\cite{guan2013fermi,chetcuti2023probe,capponi2008molecular}  can be studied for multi-component systems.

Persistent currents of multi-component systems in a ring track  have been studied in~\cite{beattie2013persistent,anoshkin2013persistent,white2017odd,spehner2021persistent,richaud2021interaction,chetcuti2022persistent,consiglio2022variational,richaud2022mimicking,chetcuti2023probe,chetcuti2023interference,osterloh2023exact,chetcuti2023persistent,pecci2023persistent}. 
Within multi-component ring-shaped lattices, it was shown that the angular momentum per particle in strongly correlated systems with SU($N$) symmetry acquires fractional values of $1/N_{p}$ ($1/N$) for repulsive~\cite{chetcuti2022persistent} (attractive~\cite{chetcuti2023probe}) fermions and of $1/N_{p}$ for repulsive bosons~\cite{pecci2023persistent} where $N_{p}$ ($N$) is the number of particles (components). Such quantization properties are expected to result in an enhanced sensitivity to rotation as predicted for attractive one-component bosons~\cite{naldesi2022enhancing,polo2020quantum}. In this context, it would be  worth exploring $N$-component matter-wave transport through  localized weak links and impurity problems~\cite{kane1992transmission,saleur1998lectures,rylands2016quantum} on mesoscopic cold atoms rings~\cite{cominotti2014optimal,cominotti2015scaling,aghamalyan2015coherent}. 
At the same time, the presence of one or more  barriers in a ring circuit provides the basis for the realization of quantum electronic-inspired devices like Atomtronic Quantum Interference Devices (AQUIDs) ~\cite{eckel2014hysteresis,ryu2020quantum,kumar2017temperature,mathey2016realizing,kiehn2022implementation,naldesi2022enhancing,aghamalyan2015atomtronics,amico2014superfluid,aghamalyanatomtronic2016,polo2020quantum,solenov2010macroscopic}. 
Besides being interesting to study questions like macroscopic phase coherence, such systems  can provide  relevant platforms also for  a matter-wave interferometry based on multi-component quantum fluids.

Digital Mirror Devices, painting techniques and Spatial Light Modulators have opened the way to study driven circuits~\cite{gauthier2016direct,rubinsztein2016roadmap,henderson2009experimental}. A single periodically driven Josephson junction in a strip of $^6$Li atoms in BEC regime is predicted to display quantized dc-ac transitions~\cite{singh2023shapiro}. Such a phenomenon is the cold atoms counterpart of the  Shapiro steps originally observed in the I-V characteristics of the superconducting Josephson junction~ \cite{shapiro1963josephson, grimes1968millimeter}. Shapiro steps in driven atomic Josephson junctions can be exploited to study emergent phenomena of superfluidity in the BEC-BCS crossover with a new twist. Analogously to their value in  metrological voltage standards~\cite{hamilton1995ieee,burroughs1999ieee,burroughs2011ieee},  
 driven Josephson junctions can define a useful tool to develop high-precision atomtronic circuitry, including more complex geometries, driven AQUIDs, or driven coupled arrays of atomic Josephson junctions.

\section{Rydberg atoms}

Rydberg atoms are atoms excited in states with a large principal quantum number. For this reason, they possess very distinctive properties including a large dipole moment that leads to a strong dipole-dipole interaction~\cite{saffman2010quantum, adams2019rydberg,browaeys2020many, wu2021concise,morgado2021quantum}. 
Atoms residing in the same Rydberg state effectively interact through a
van der Waals-like potential with a characteristic $1/R^6$, with $R$ being the atom-atom spatial distance ~\cite{bernien2017probing,pohl2010dynamical}. Instead, the dipole-dipole interactions of atoms in Rydberg states of opposite parity result in a spin exchange coupling that scales as $1/R^3$~\cite{barredo2015coherent}.
The coherent local addressability and manipulation of Rydberg atoms in a large variety of spatial configurations have been thoroughly demonstrated ~\cite{endres2016atom,barredo2018synthetic,barredo2016atom,schymik2020enhanced}.
A spectacular phenomenon arising from the strong interaction is the dipole blockade for which only one atom in a characteristic spatial radius can be excited to a Rydberg state~\cite{lukin2001dipole, urban2009observation}. Moreover, by shining on the systems with a suitable laser field, a specific detuning on the atomic energy levels can be induced, which ultimately results in a facilitation (or anti-blockade) mechanism
~\cite{valado2016experimental,morsch2018many,lesanovsky2014out}. Controlling these blockade and anti-blockade protocols in networks of Rydberg atoms can provide a very versatile platform for quantum simulation~\cite{adams2019rydberg,browaeys2020many,morsch2018many}.

Rydberg atoms can indeed define a new concept for atomtronics in that the flowing current in the circuits, rather than matter, can occur in terms of Rydberg excitations. Compared with the millisecond dynamics of cold atoms, Rydberg excitations can propagate on a microsecond time scales. As a result, faster atomtronic circuits and devices can be achieved.

Networks of Rydberg atoms are a promising platform to realize quantum state transfer, with multiple protocols being widely theoretically studied in quantum spin systems~\cite{bose2003quantum,christandl2004perfect,paganelli2013routing,apollaro2015many,chetcuti2020perturbative, apollaro2020two, apollaro2022quantum} and some being carried over with Rydberg atoms~\cite{palaiodimopulos2023chiral,li2019shao}. The remarkable control on the parameters of the system, like the aforementioned local addressability obtained with optical tweezers~\cite{barredo2016atom, endres2016atom, moffitt2008recent} and interactions, gives the possibility to conceive atomtronic devices to transfer entangled and single excitation states. The basis to realize iconic quantum transport devices in mesoscopic physics like rf- and dc-SQUIDs has been proposed~\cite{perciavalle2023coherent}. An artificial gauge flux which results in a controllable flow of excitation in rings has been theoretically studied with different techniques~\cite{wu2022manipulating,perciavalle2023controlled} and experimentally realized for small systems~\cite{lienhard2020realization}. Putting together these two features and considering the flexibility of the geometry, Rydberg atoms can constitute a promising platform for the realization of interference devices based on excitation currents,
they can also be used to propose and implement source-drain quantum transport with new and exotic channel configurations. 

By suitable engineering of the dipole anti-blockade, Rydberg atomtronic circuits emulating electronic devices, such as a diode, switches and logical gates have been proposed~\cite{kitson2023rydberg}. With this set of devices, the development of more intricate gadgets and more complex circuits can be proposed, examples may include adders and routers, which can be conceived through the implementation of the universal logical gate set. 
In addition, the manipulation of dissipation~\cite{begoc2023controlled} also defines a further knob to manipulate the Rydberg atomtronic circuits dynamically, which can be used for conceiving new devices as well as devising quantum simulation of driven-dissipative systems~\cite{marcuzzi2016absorbing,morsch2018many,wintermantel2021epidemic}.

Another promising pathway of research, that exploits the potential and control offered by Rydberg atom platforms, is the quantum simulation of high-energy physics. This field has recently attracted the interest of a very heterogeneous community, aimed at exploiting quantum technologies to study physical phenomena, such as confinement, which are elusive by nature. In this context, digital and quantum simulations of lattice gauge theories (LGTs) have been widely explored in several platforms, ranging from ultracold atoms in optical lattices~\cite{zohar2016quantum,mil2020scalable,yang2020observation,zhao2022thermalization,halimeh2023cold,surace2023ab,surace2023scalable,schweizer2019floquet} and Rydberg atoms~\cite{weimer2010rydberg,celi2020emerging,surace2020lattice}, to trapped ions~\cite{davoudi2020towards,buazuavan2023synthetic,nguyen2022digital} and superconducting circuits~\cite{atas2021su2,atas2023simulating,wang2022observation,mildenberger2022probing}. Rydberg atoms, which interact according to a Van der Waals-like behaviour, naturally implement the gauge-invariant dynamics of fermionic matter coupled to $\mathbb{Z}_2$ gauge fields in a specific gauge sector~\cite{surace2021scattering}. These allow for the study of coherence properties of confined matter, such as the peculiar Aharonov-Bohm oscillations that emerge, at the mesoscopic scale, in the presence of an external magnetic flux~\cite{domanti2023coherence}. Such magnetic flux could be experimentally implemented through Floquet driving schemes applied to the detuning of the atoms, thus realizing a coherent current of confined matter. 
In these cases, the control over the experimental parameters in a Rydberg atom platform is instrumental. Indeed, the time scales required for the observation of the effective gauge invariant real-time dynamics are within reach of the currently attained experimental times.

\section{Hybrid devices}

As quantum technologies continue to advance, the seamless integration of atomtronic circuits with established technologies becomes increasingly important. This requires capitalizing on the unique features of different quantum systems to construct hybrid quantum circuits with enhanced capabilities compared with classical counterparts, thus enabling to unlock new technologies.
Examples of hybrid systems based on ultracold atoms include the use of cavities \cite{kumar2021cavity,pradhan2023cavity} as well as superconducting circuits disks \cite{zhang2019magnetic}, films \cite{muller2010trapping}, wires \cite{cano2009meissner} and rings \cite{mukai2007persistent}. 

Many hybrid systems stem from the extensive overlap between distinctive fields. A prime example is the fusion of quantum optics and ultracold matter. These two domains have been mutually beneficial ever since the inception of ultracold matter. 
An example of such a hybridization can be found in the use of optical nanofibers surrounded by a cloud of cold atoms, where the interaction between light and matter is facilitated by the evanescent field generated by the nanofiber \cite{chormaic2023probing,russell2013laser,goban2012demonstration,sague2007cold} which can also be used to create spatially dependent artificial magnetic fields with new and exciting phenomena~\cite{westbrook1998new,gillen2009twodimensional,schloss2020controlled,hejazi2020symmetry,hejazi2022formation}. 
Extending hybrid circuits of ultracold atoms coupled with superconductors can be very relevant for quantum technologies. The general idea behind is that in a schematic
information processing protocol, gate operations and  state preparations can be carried out in the fast solid-state apparatus; then transferred and stored in
atomic systems that are less prone to decoherence  and finally transferred  back to solid-state devices for
further processing. Relying on the sensitivity of the atomic spins to microwaves~\cite{fleischhauer2005electromagnetically,hammerer2010quantum}, coupled systems of atoms and superconducting elements have been proposed~\cite{xiang_hybrid_2013,yu2016charge,yu2017superconducting,yu2016superconducting,yu2016quantum,yu2018stabilizing,yu2017theoretical,yu2018charge}.
The fabrication of various magnetic traps for ultracold atoms on superconducting atom
chips~\cite{cano2008meissner,mukai2007persistent,nirrengarten2006realization,muller2010trapping,tosto2019optically,muller2010programmable,PhysRevA.85.013404} together with the studies in quantum information processing on coupling ultracold atoms and superconducting chips~\cite{verdu2009strong,petrosyan2019microwave,bernon2013manipulation,
hattermann2017coupling, fortagh2007magnetic} paves the
way towards the realization of an atomic quantum memory linked to superconducting quantum circuits.

More recently, the combination of the large coherence times of ultracold matter with opto-mechanics has been proposed as a hybrid system that can provide much higher sensitivity and control of the atoms. In particular, this technology can allow us to probe weakly interacting bosonic atoms in a non-destructive way \cite{kumar2021cavity} as well as probing more sensitive regimes such as attractively interacting bosonic systems that can form solitonic states with application to interferometry \cite{pradhan2023cavity}. Remarkably, due to the long coherence times provided by the cold atoms, their use in the context of quantum memories has been considered for integrated quantum computation. 
In particular, these systems provide advantages in both coherence times and high retrieval rates by making use of the aforementioned control on the light-matter interaction \cite{bao2012efficient,cho2016coherent}. 

Expanding these hybridizations between atomtronics and quantum technologies is especially important in the fields of sensing and simulation. By utilizing the unique characteristics of different platforms, such as the ones discussed in this perspective, we can significantly broaden the scope of the field, unlocking the full potential of quantum systems 
and promoting progress in this rapidly evolving field.


\section{Conclusions}

In this article, we highlight how new platforms in cold atoms quantum matter, other than the simple bosonic or fermionic degenerate gases, 
can open up new research opportunities in atomtronics-enabled quantum technologies. 
We focused on three specific platforms  defining research  directions with great potential: multi-component gases, Rydberg atoms and hybrid systems.
Key areas that can benefit from these are matter-wave sensing technology and quantum simulations with reduced time scales.

To close this article we comment on the important role that machine learning can have on the design and control of the future atomtronic quantum technologies. Indeed,  through machine learning optimization has been already implemented in  specific experimental schemes of ultracold bosonic systems as  cooling processes \cite{wigley2016fast,barker2020applying}  read-out protocols~\cite{ness2020single,metz2021deep,kim2023vortex}. 
The new implementations discussed in this perspective can also profit from these techniques.
In particular, Reinforcement Learning (RL) where one can analyze complex systems without the need of prior knowledge, can provide a practical solution for the system's experimental control problems without relying on assumptions about system and manual analysis.
Recent examples in experimental atomtronics circuits include the use of RL for optimizing matter-wave interferometer \cite{chih2021reinforcement,abad2014persistent} and for imparting  persistent currents in ring geometries\cite{simjanovski2023optimizing,haug2019machine}. 
Nonetheless, the possibilities offered by machine learning in atomtronics circuits are ample, and a vast territory is unexplored. Further developing proof of concept devices that use machine learning as a tool to drive quantum matter, being either multi-component gases or Rydberg, would broaden the field; on the other way around, atomtronics-inspired machine learning schemes can find applications in other quantum technologies ~\cite{chih2021reinforcement,abad2014persistent}.
Finally, quantum technologies on the cloud, such as Infleqtion (ColdQuanta) for cold atoms~\cite{north2023albert} or QuEra for Rydberg atoms \cite{huber2022cloud}, can open new important avenues for the field.

\ack{PK and LA acknowledge the Julian Schwinger Foundation grant JSF-18-12-0011.}

\bibliographystyle{iopart-num.bst}
\bibliography{ref}

\providecommand{\newblock}{}
\begin{thebibliography}{100}
\expandafter\ifx\csname url\endcsname\relax
  \def\url#1{{\tt #1}}\fi
\expandafter\ifx\csname urlprefix\endcsname\relax\def\urlprefix{URL }\fi
\providecommand{\eprint}[2][]{\url{#2}}

\bibitem{gauthier2016direct}
Gauthier G, Lenton I, Parry N~M, Baker M, Davis M~J, Rubinsztein-Dunlop H and
  Neely T~W 2016 {\em Optica\/} {\bf 3} 1136--1143
  \urlprefix\url{https://opg.optica.org/optica/abstract.cfm?URI=optica-3-10-1136}

\bibitem{rubinsztein2016roadmap}
Rubinsztein-Dunlop H, Forbes A, Berry M~V, Dennis M~R, Andrews D~L, Mansuripur
  M {\em et~al.\/} 2016 {\em Journal of Optics\/} {\bf 19} 013001
  \urlprefix\url{https://dx.doi.org/10.1088/2040-8978/19/1/013001}

\bibitem{henderson2009experimental}
Henderson K, Ryu C, MacCormick C and Boshier M~G 2009 {\em New Journal of
  Physics\/} {\bf 11} 043030 ISSN 1367-2630
  \urlprefix\url{http://dx.doi.org/10.1088/1367-2630/11/4/043030}

\bibitem{bloch2005ultracold}
Bloch I 2005 {\em Nature Physics\/} {\bf 1} 23–30 ISSN 1745-2481
  \urlprefix\url{http://dx.doi.org/10.1038/nphys138}

\bibitem{cornell2002nobel}
Cornell E~A and Wieman C~E 2002 {\em Reviews of Modern Physics\/} {\bf 74} 875
  \urlprefix\url{https://journals.aps.org/rmp/pdf/10.1103/RevModPhys.74.875}

\bibitem{ketterle2002nobel}
Ketterle W 2002 {\em Reviews of Modern Physics\/} {\bf 74} 1131
  \urlprefix\url{https://journals.aps.org/rmp/pdf/10.1103/RevModPhys.74.1131}

\bibitem{dalfovo1999theory}
Dalfovo F, Giorgini S, Pitaevskii L~P and Stringari S 1999 {\em Reviews of
  Modern Physics\/} {\bf 71}(3) 463--512
  \urlprefix\url{https://link.aps.org/doi/10.1103/RevModPhys.71.463}

\bibitem{seaman2007atomtronics}
Seaman B~T, Kr\"amer M, Anderson D~Z and Holland M~J 2007 {\em Physical Review
  A\/} {\bf 75}(2) 023615
  \urlprefix\url{https://link.aps.org/doi/10.1103/PhysRevA.75.023615}

\bibitem{amico2017focus}
Amico L, Birkl G, Boshier M and Kwek L~C 2017 {\em New Journal of Physics\/}
  {\bf 19} 020201 \urlprefix\url{https://dx.doi.org/10.1088/1367-2630/aa5a6d}

\bibitem{amico2021roadmap}
Amico L, Boshier M, Birkl G, Minguzzi A, Miniatura C, Kwek L~C {\em et~al.\/}
  2021 {\em AVS Quantum Science\/} {\bf 3} 039201 ISSN 2639-0213
  (\textit{Preprint}
  \eprint{https://pubs.aip.org/avs/aqs/article-pdf/doi/10.1116/5.0026178/16662758/039201\_1\_online.pdf})
  \urlprefix\url{https://doi.org/10.1116/5.0026178}

\bibitem{amico2022colloquium}
Amico L, Anderson D, Boshier M, Brantut J~P, Kwek L~C, Minguzzi A and von
  Klitzing W 2022 {\em Reviews of Modern Physics\/} {\bf 94}(4) 041001
  \urlprefix\url{https://link.aps.org/doi/10.1103/RevModPhys.94.041001}

\bibitem{dekker2000guiding}
Dekker N~H, Lee C~S, Lorent V, Thywissen J~H, Smith S~P, Drndi{\'c} M,
  Westervelt R~M and Prentiss M 2000 {\em Physical Review Letters\/} {\bf 84}
  1124--1127 \urlprefix\url{https://dx.doi.org/10.1103/PhysRevLett.84.1124}

\bibitem{dumke2002interferometer}
Dumke R, M\"{u}ther T, Volk M, Ertmer W and Birkl G 2002 {\em Physical Review
  Letters\/} {\bf 89} 220402
  \urlprefix\url{https://doi.org/10.1103/PhysRevLett.89.220402}

\bibitem{leanhardt2002propagation}
Leanhardt A, Chikkatur A, Kielpinski D, Shin Y, Gustavson T, Ketterle W and
  Pritchard D 2002 {\em Physical Review Letters\/} {\bf 89} 040401
  \urlprefix\url{https://doi.org/10.1103/PhysRevLett.89.040401}

\bibitem{muller2000guiding}
M\"{u}ller D, Cornell E~A, Anderson D~Z and Abraham E~R~I 2000 {\em Physical
  Review A\/} {\bf 61} 033411
  \urlprefix\url{https://doi.org/10.1103/PhysRevA.61.033411}

\bibitem{schneble2003integrated}
Schneble D, Hasuo M, Anker T, Pfau T and Mlynek J 2003 {\em Journal of the
  Optical Society of America B\/} {\bf 20} 648--651
  \urlprefix\url{https://doi.org/10.1364/JOSAB.20.000648}

\bibitem{denschlag1999guiding}
Denschlag J, Cassettari D and Schmiedmayer J 1999 {\em Physical Review
  Letters\/} {\bf 82} 2014 \urlprefix\url{10.1103/PhysRevLett.82.2014}

\bibitem{denschlag1999neutral}
Denschlag J, Cassettari D, Chenet A, Schneider S and Schmiedmayer J 1999 {\em
  Applied Physics B\/} {\bf 69} 291--301
  \urlprefix\url{https://doi.org/10.1007/s003400050809}

\bibitem{schmiedmayer1995wire}
Schmiedmayer J 1995 {\em Applied Physics B\/} {\bf 60} 169--179
  \urlprefix\url{https://doi.org/10.1007/BF01135859}

\bibitem{schmiedmayer1996classical}
Schmiedmayer J and Scrinzi A 1996 {\em Quantum and Semiclassical Optics:
  Journal of the European Optical Society Part B\/} {\bf 8} 693
  \urlprefix\url{https://iopscience.iop.org/article/10.1088/1355-5111/8/3/029/pdf}

\bibitem{schmiedmayer1996particle}
Schmiedmayer J and Scrinzi A 1996 {\em Physical Review A\/} {\bf 54} R2525
  \urlprefix\url{https://doi.org/10.1103/PhysRevA.54.R2525}

\bibitem{schmiedmayer1995guiding}
Schmiedmayer J 1995 {\em Physical Review A\/} {\bf 52} R13--R16
  \urlprefix\url{https://doi.org/10.1103/PhysRevA.52.R13}

\bibitem{pepino2021entropy}
Pepino R~A 2021 {\em Entropy\/} {\bf 23} ISSN 1099-4300
  \urlprefix\url{https://www.mdpi.com/1099-4300/23/5/534}

\bibitem{pepino2009atomtronic}
Pepino R~A, Cooper J, Anderson D~Z and Holland M~J 2009 {\em Physical Review
  Letters\/} {\bf 103}(14) 140405
  \urlprefix\url{https://link.aps.org/doi/10.1103/PhysRevLett.103.140405}

\bibitem{gou2020tunable}
Gou W, Chen T, Xie D, Xiao T, Deng T~S, Gadway B, Yi W and Yan B 2020 {\em
  Physical Review Letters\/} {\bf 124}(7) 070402
  \urlprefix\url{https://link.aps.org/doi/10.1103/PhysRevLett.124.070402}

\bibitem{stickney2007transistorlike}
Stickney J~A, Anderson D~Z and Zozulya A~A 2007 {\em Physical Review A\/} {\bf
  75}(1) 013608
  \urlprefix\url{https://link.aps.org/doi/10.1103/PhysRevA.75.013608}

\bibitem{caliga2016principles}
Caliga S~C, Straatsma C~J~E, Zozulya A~A and Anderson D~Z 2016 {\em New Journal
  of Physics\/} {\bf 18} 015012
  \urlprefix\url{https://dx.doi.org/10.1088/1367-2630/18/1/015012}

\bibitem{mclain2018quantum}
McLain M~A and Carr L~D 2018 {\em Quantum Science and Technology\/} {\bf 3}
  035012 \urlprefix\url{https://dx.doi.org/10.1088/2058-9565/aac731}

\bibitem{wilsmann2018control}
Wilsmann K~W, Ymai L~H, Tonel A~P, Links J and Foerster A 2018 {\em
  Communications Physics\/} {\bf 1}
  \urlprefix\url{https://doi.org/10.1038/s42005-018-0089-1}

\bibitem{zozulya2013principles}
Zozulya A~A and Anderson D~Z 2013 {\em Physical Review A\/} {\bf 88}(4) 043641
  \urlprefix\url{https://link.aps.org/doi/10.1103/PhysRevA.88.043641}

\bibitem{caliga2017experimental}
Caliga S~C, Straatsma C~J~E and Anderson D~Z 2017 {\em New Journal of
  Physics\/} {\bf 19} 013036
  \urlprefix\url{https://dx.doi.org/10.1088/1367-2630/aa56d8}

\bibitem{caliga2012matterwave}
Caliga S~C, Straatsma C~J~E, Zozulya A~A and Anderson D~Z 2012 {\em ArXiv
  e-prints\/}  1208.3109
  \urlprefix\url{https://doi.org/10.48550/arXiv.1208.3109}

\bibitem{caliga2016transport}
Caliga S~C, Straatsma C~J and Anderson D~Z 2016 {\em New Journal of Physics\/}
  {\bf 18} 025010
  \urlprefix\url{https://iopscience.iop.org/article/10.1088/1367-2630/18/2/025010}

\bibitem{amico2005quantum}
Amico L, Osterloh A and Cataliotti F 2005 {\em Physical Review Letters\/} {\bf
  95} 063201 \urlprefix\url{https://doi.org/10.1103/PhysRevLett.95.063201}

\bibitem{wright2013driving}
Wright K~C, Blakestad R~B, Lobb C~J, Phillips W~D and Campbell G~K 2013 {\em
  Physical Review Letters\/} {\bf 110}(2) 025302
  \urlprefix\url{https://link.aps.org/doi/10.1103/PhysRevLett.110.025302}

\bibitem{aghamalyan2015coherent}
Aghamalyan D, Cominotti M, Rizzi M, Rossini D, Hekking F, Minguzzi A, Kwek L~C
  and Amico L 2015 {\em New Journal of Physics\/} {\bf 17} 045023
  \urlprefix\url{https://dx.doi.org/10.1088/1367-2630/17/4/045023}

\bibitem{haug2018readout}
Haug T, Tan J, Theng M, Dumke R, Kwek L~C and Amico L 2018 {\em Physical Review
  A\/} {\bf 97}(1) 013633
  \urlprefix\url{https://link.aps.org/doi/10.1103/PhysRevA.97.013633}

\bibitem{ryu2020quantum}
Ryu C, Samson E~C and Boshier M~G 2020 {\em Nature Communications\/} {\bf 11}
  \urlprefix\url{https://doi.org/10.1038/s41467-020-17185-6}

\bibitem{pezze2023stabilizing}
Pezzè L, Xhani K, Daix C, Grani N, Donelli B, Scazza F, Hernandez-Rajkov D,
  Kwon W~J, Pace G~D and Roati G 2023 {Stabilizing persistent currents in an
  atomtronic Josephson junction necklace} (\textit{Preprint}
  \eprint{2311.05523})
  \urlprefix\url{https://doi.org/10.48550/arXiv.2311.05523}

\bibitem{wang2015resonant}
Wang Y~H, Kumar A, Jendrzejewski F, Wilson R~M, Edwards M, Eckel S, Campbell
  G~K and Clark C~W 2015 {\em New Journal of Physics\/} {\bf 17} 125012
  \urlprefix\url{https://dx.doi.org/10.1088/1367-2630/17/12/125012}

\bibitem{matthew2015self}
Mathew R, Kumar A, Eckel S, Jendrzejewski F, Campbell G~K, Edwards M and
  Tiesinga E 2015 {\em Physical Review A\/} {\bf 92} 033602
  \urlprefix\url{https://doi.org/10.1103/PhysRevA.92.033602}

\bibitem{yakimenko2015vortices}
Yakimenko A, Bidasyuk Y, Weyrauch M, Kuriatnikov Y and Vilchinskii S 2015 {\em
  Physical Review A\/} {\bf 91} 033607
  \urlprefix\url{https://doi.org/10.1103/PhysRevA.91.033607}

\bibitem{mehdi2021superflow}
Mehdi Z, Bradley A~S, Hope J~J and Szigeti S~S 2021 {\em SciPost Physics\/}
  {\bf 11}(4) 80
  \urlprefix\url{https://scipost.org/10.21468/SciPostPhys.11.4.080}

\bibitem{mathey2016realizing}
Mathey A~C and Mathey L 2016 {\em New Journal of Physics\/} {\bf 18} 055016
  \urlprefix\url{https://dx.doi.org/10.1088/1367-2630/18/5/055016}

\bibitem{rastelli2013quantum}
Rastelli G, Pop I~M and Hekking F~W 2013 {\em Physical Review B\/} {\bf 87}
  174513 \urlprefix\url{https://doi.org/10.1103/PhysRevB.87.174513}

\bibitem{schenke2011nonadiabatic}
Schenke C, Minguzzi A and Hekking F~W~J 2011 {\em Physical Review A\/} {\bf 84}
  053636 \urlprefix\url{https://doi.org/10.1103/PhysRevA.84.053636}

\bibitem{amico2014superfluid}
Amico L, Aghamalyan D, Auksztol F, Crepaz H, Dumke R and Kwek L~C 2014 {\em
  Scientific Reports\/} {\bf 4}
  \urlprefix\url{https://doi.org/10.1038/srep04298}

\bibitem{solenov2010macroscopic}
Solenov D and Mozyrsky D 2010 {\em Physical Review A\/} {\bf 82} 061601
  \urlprefix\url{https://doi.org/10.1103/PhysRevA.82.061601}

\bibitem{hallwood2006macroscopic}
Hallwood D~W, Burnett K and Dunningham J 2006 {\em New Journal of Physics\/}
  {\bf 8} 180
  \urlprefix\url{https://iopscience.iop.org/article/10.1088/1367-2630/8/9/180/pdf}

\bibitem{nunnenkamp2008generation}
Nunnenkamp A, Rey A~M and Burnett K 2008 {\em Physical Review A\/} {\bf 77}
  023622 \urlprefix\url{https://doi.org/10.1103/PhysRevA.77.023622}

\bibitem{safaei2018two}
Safaei S, Gr\'{e}maud B, Dumke R, Kwek L~C, Amico L and Miniatura C 2018 {\em
  Physical Review A\/} {\bf 97} 042306
  \urlprefix\url{https://doi.org/10.1103/PhysRevA.97.042306}

\bibitem{bloch2008many}
Bloch I, Dalibard J and Zwerger W 2008 {\em Reviews of Modern Physics\/} {\bf
  80} 885--964 \urlprefix\url{https://doi.org/10.1103/RevModPhys.80.885}

\bibitem{buluta2009quantum}
Buluta I and Nori F 2009 {\em Science\/} {\bf 326} 108--111
  \urlprefix\url{https://doi.org/10.1126/science.117783}

\bibitem{cirac2012goals}
Cirac J~I and Zoller P 2012 {\em Nature Physics\/} {\bf 8} 264--266
  \urlprefix\url{https://doi.org/10.1038/nphys2275}

\bibitem{dowling2003quantum}
Dowling J~P and Milburn G~J 2003 {\em Philosophical Transactions of the Royal
  Society of London A\/} {\bf 361} 1655--1674
  \urlprefix\url{https://doi.org/10.1098/rsta.2003.1227}

\bibitem{lamata2014epj}
Lamata L, Mezzacapo A, Casanova J, Solano E, Johnson T~H, Clark S~R, Jaksch D,
  Krutitsky K~V, Navez P, Queisser F {\em et~al.\/} 2014 {\em Quantum\/} {\bf
  1}
  \urlprefix\url{https://link.springer.com/journal/40507/volumes-and-issues/1-1}

\bibitem{lewenstein2012ultracold}
Lewenstein M, Sanpera A and Ahufinger V 2012 {\em {Ultracold Atoms in Optical
  Lattices: Simulating quantum many-body systems}\/} (Oxford University Press)
  \urlprefix\url{https://doi.org/10.1093/acprof:oso/9780199573127.001.0001}

\bibitem{moulder2012quantized}
Moulder S, Beattie S, Smith R~P, Tammuz N and Hadzibabic Z 2012 {\em Physical
  Review A\/} {\bf 86}(1) 013629
  \urlprefix\url{https://link.aps.org/doi/10.1103/PhysRevA.86.013629}

\bibitem{pandey2019hypersonic}
Pandey S, Mas H, Drougakis G, Thekkeppatt P, Bolpasi V, Vasilakis G, Poulios K
  and von Klitzing W 2019 {\em Nature\/} {\bf 570} 1
  \urlprefix\url{https://doi.org/10.1038/s41586-019-1273-5}

\bibitem{guo2020supersonic}
Guo Y, Dubessy R, de~Herve M~d~G, Kumar A, Badr T, Perrin A, Longchambon L and
  Perrin H 2020 {\em Physical Review Letters\/} {\bf 124} 025301
  \urlprefix\url{https://doi.org/10.1103/PhysRevLett.124.025301}

\bibitem{pace2022imprinting}
Del~Pace G, Xhani K, Muzi~Falconi A, Fedrizzi M, Grani N, Hernandez~Rajkov D,
  Inguscio M, Scazza F, Kwon W~J and Roati G 2022 {\em Physical Review X\/}
  {\bf 12}(4) 041037
  \urlprefix\url{https://link.aps.org/doi/10.1103/PhysRevX.12.041037}

\bibitem{cai2022persistent}
Cai Y, Allman D~G, Sabharwal P and Wright K~C 2022 {\em Physical Review
  Letters\/} {\bf 128}(15) 150401
  \urlprefix\url{https://link.aps.org/doi/10.1103/PhysRevLett.128.150401}

\bibitem{kumar2017temperature}
Kumar A, Eckel S, Jendrzejewski F and Campbell G~K 2017 {\em Physical Review
  A\/} {\bf 95} 021602
  \urlprefix\url{https://doi.org/10.1103/PhysRevA.95.021602}

\bibitem{jendrzejewski2014resistive}
Jendrzejewski F, Eckel S, Murray N, Lanier C, Edwards M, Lobb C~J and Campbell
  G~K 2014 {\em Physical Review Letters\/} {\bf 113} 045305
  \urlprefix\url{https://doi.org/10.1103/PhysRevLett.113.045305}

\bibitem{eckel2014hysteresis}
Eckel S, Lee J~G, Jendrzejewski F, Murray N, Clark C~W, Lobb C~J, Phillips W~D,
  Edwards M and Campbell G~K 2014 {\em Nature\/} {\bf 506} 200{\textendash}203

\bibitem{cominotti2014optimal}
Cominotti M, Rossini D, Rizzi M, Hekking F and Minguzzi A 2014 {\em Physical
  Review Letters\/} {\bf 113} 025301
  \urlprefix\url{https://doi.org/10.1103/PhysRevLett.113.025301}

\bibitem{hekking1997quantum}
Hekking F and Glazman L 1997 {\em Physical Review B\/} {\bf 55} 6551
  \urlprefix\url{https://doi.org/10.1103/PhysRevB.55.6551}

\bibitem{matveev2002persistent}
Matveev K, Larkin A and Glazman L 2002 {\em Physical Review Letters\/} {\bf 89}
  096802 \urlprefix\url{https://doi.org/10.1103/PhysRevLett.89.096802}

\bibitem{haug2019machine}
Haug T, Dumke R, Kwek L~C, Miniatura C and Amico L 2021 {\em Physical Review
  Research\/} {\bf 3}(1) 013034
  \urlprefix\url{https://doi.org/10.1103/PhysRevResearch.3.013034}

\bibitem{aghamalyan2013effective}
Aghamalyan D, Amico L and Kwek L~C 2013 {\em Physical Review A\/} {\bf 88}(6)
  063627 \urlprefix\url{https://link.aps.org/doi/10.1103/PhysRevA.88.063627}

\bibitem{haug2018mesoscopic}
Haug T, Amico L, Dumke R and Kwek L~C 2018 {\em Quantum Science and
  Technology\/} {\bf 3} 035006
  \urlprefix\url{https://dx.doi.org/10.1088/2058-9565/aaa8c6}

\bibitem{pecci2021probing}
Pecci G, Naldesi P, Amico L and Minguzzi A 2021 {\em Physical Review
  Research\/} {\bf 3}(3) L032064
  \urlprefix\url{https://link.aps.org/doi/10.1103/PhysRevResearch.3.L032064}

\bibitem{pecci2023single}
Pecci G, Naldesi P, Minguzzi A and Amico L 2022 {\em Quantum Science and
  Technology\/} {\bf 8} 01LT03
  \urlprefix\url{https://dx.doi.org/10.1088/2058-9565/aca712}

\bibitem{polo2016transport}
Polo J, Benseny A, Busch T, Ahufinger V and Mompart J 2016 {\em New Journal of
  Physics\/} {\bf 18} 015010
  \urlprefix\url{https://dx.doi.org/10.1088/1367-2630/18/1/015010}

\bibitem{richaud2017quantum}
Richaud A and Penna V 2017 {\em Physical Review A\/} {\bf 96}(1) 013620
  \urlprefix\url{https://link.aps.org/doi/10.1103/PhysRevA.96.013620}

\bibitem{polo2018damping}
Polo J, Ahufinger V, Hekking F~W~J and Minguzzi A 2018 {\em Physical Review
  Letters\/} {\bf 121}(9) 090404
  \urlprefix\url{https://link.aps.org/doi/10.1103/PhysRevLett.121.090404}

\bibitem{victorin2019nonclassical}
Victorin N, Haug T, Kwek L~C, Amico L and Minguzzi A 2019 {\em Physical Review
  A\/} {\bf 99} 033616
  \urlprefix\url{https://doi.org/10.1103/PhysRevA.99.033616}

\bibitem{lesanovsky2007time}
Lesanovsky I and von Klitzing W 2007 {\em Physical Review Letters\/} {\bf 99}
  083001 \urlprefix\url{https://doi.org/10.1103/PhysRevLett.99.083001}

\bibitem{nicolau2020orbital}
Nicolau E, Mompart J, Juli\'a-D\'{\i}az B and Ahufinger V 2020 {\em Physical
  Review A\/} {\bf 102} 023331
  \urlprefix\url{https://doi.org/10.1103/PhysRevA.102.023331}

\bibitem{oliinyk2019tunneling}
Oliinyk A, Yakimenko A and Malomed B 2019 {\em Journal of Physics B: Atomic,
  Molecular and Optical Physics\/} {\bf 52} 225301
  \urlprefix\url{https://dx.doi.org/10.1088/1361-6455/ab46f9}

\bibitem{danshita2012quantum}
Danshita I and Polkovnikov A 2012 {\em Physical Review A\/} {\bf 85} 023638
  \urlprefix\url{https://doi.org/10.1103/PhysRevA.85.023638}

\bibitem{kunimi2017thermally}
Kunimi M and Danshita I 2017 {\em Physical Review A\/} {\bf 95} 033637
  \urlprefix\url{https://doi.org/10.1103/PhysRevA.95.033637}

\bibitem{eller2020producing}
Eller B, Oladehin O, Fogarty D, Heller C, Clark C~W and Edwards M 2020 {\em
  Physical Review A\/} {\bf 102} 063324
  \urlprefix\url{https://doi.org/10.1103/PhysRevA.102.063324}

\bibitem{murray2013probing}
Murray N, Krygier M, Edwards M, Wright K~C, Campbell G~K and Clark C~W 2013
  {\em Physical Review A\/} {\bf 88} 053615
  \urlprefix\url{10.1103/PhysRevA.88.053615}

\bibitem{corman2014quench}
Corman L, Chomaz L, Bienaim\'{e} T, Desbuquois R, Weitenberg C, Nascimbene S,
  Dalibard J and Beugnon J 2014 {\em Physical Review Letters\/} {\bf 113}
  135302 \urlprefix\url{https://doi.org/10.1103/PhysRevLett.113.135302}

\bibitem{roscilde2016from}
Roscilde T, Faulkner M~F, Bramwell S~T and Holdsworth P~C~W 2016 {\em New
  Journal of Physics\/} {\bf 18} 075003
  \urlprefix\url{https://iopscience.iop.org/article/10.1088/1367-2630/18/7/075003}

\bibitem{waintal2008persistent}
Waintal X, Fleury G, Kazymyrenko K, Houzet M, Schmitteckert P and Weinmann D
  2008 {\em Physical Review Letters\/} {\bf 101} 106804
  \urlprefix\url{https://doi.org/10.1103/PhysRevLett.101.106804}

\bibitem{yu1992persistent}
Yu N and Fowler M 1992 {\em Physical Review B\/} {\bf 45} 11795--11804
  \urlprefix\url{https://doi.org/10.1103/PhysRevB.45.11795}

\bibitem{patu2022temperature}
P{\^a}{\c{t}}u O~I and Averin D~V 2022 {\em Physical Review Letters\/} {\bf
  128} 096801 \urlprefix\url{https://doi.org/10.1103/PhysRevLett.128.096801}

\bibitem{arabahmadi2021universal}
Arabahmadi E, Schumayer D and Hutchinson D~A~W 2021 {\em Physical Review A\/}
  {\bf 103} 043319 \urlprefix\url{https://doi.org/10.1103/PhysRevA.103.043319}

\bibitem{dubessy2012critical}
Dubessy R, Liennard T, Pedri P and Perrin H 2012 {\em Physical Review A\/} {\bf
  86} 011602

\bibitem{wright2013threshold}
Wright K~C, Blakestad R~B, Lobb C~J, Phillips W~D and Campbell G~K 2013 {\em
  Physical Review A\/} {\bf 88} 063633
  \urlprefix\url{https://doi.org/10.1103/PhysRevA.88.063633}

\bibitem{mathey2014decay}
Mathey A~C, Clark C~W and Mathey L 2014 {\em Physical Review A\/} {\bf 90}
  023604 \urlprefix\url{https://doi.org/10.1103/PhysRevA.90.023604}

\bibitem{abad2016persistent}
Abad M 2016 {\em Physical Review A\/} {\bf 93}(3) 033603
  \urlprefix\url{https://doi.org/10.1103/PhysRevA.93.033603}

\bibitem{burchianti2018connecting}
Burchianti A, Scazza F, Amico A, Valtolina G, Seman J~A, Fort C, Zaccanti M,
  Inguscio M and Roati G 2018 {\em Phys. Rev. Lett.\/} {\bf 120}(2) 025302
  \urlprefix\url{https://link.aps.org/doi/10.1103/PhysRevLett.120.025302}

\bibitem{polo2019oscillations}
Polo J, Dubessy R, Pedri P, Perrin H and Minguzzi A 2019 {\em Physical Review
  Letters\/} {\bf 123} 195301
  \urlprefix\url{https://doi.org/10.1103/PhysRevLett.123.195301}

\bibitem{naldesi2019rise}
Naldesi P, Gomez J~P, Malomed B, Olshanii M, Minguzzi A and Amico L 2019 {\em
  Physical Review Letters\/} {\bf 122} 053001
  \urlprefix\url{https://doi.org/10.1103/PhysRevLett.122.053001}

\bibitem{xhani2020critical}
Xhani K, Neri E, Galantucci L, Scazza F, Burchianti A, Lee K~L, Barenghi C~F,
  Trombettoni A, Inguscio M, Zaccanti M, Roati G and Proukakis N~P 2020 {\em
  Physical Review Letters\/} {\bf 124} 045301
  \urlprefix\url{https://doi.org/10.1103/PhysRevLett.124.045301}

\bibitem{naldesi2022enhancing}
Naldesi P, Polo J, Dunjko V, Perrin H, Olshanii M, Amico L and Minguzzi A 2022
  {\em SciPost Physics\/} {\bf 12} 138
  \urlprefix\url{https://scipost.org/10.21468/SciPostPhys.12.4.138}

\bibitem{perezobiol2022coherent}
P\'erez-Obiol A, Polo J and Amico L 2022 {\em Phys. Rev. Res.\/} {\bf 4}(2)
  L022038
  \urlprefix\url{https://link.aps.org/doi/10.1103/PhysRevResearch.4.L022038}

\bibitem{singh2020josephson}
Singh V~P, Luick N, Sobirey L and Mathey L 2020 {\em Phys. Rev. Res.\/} {\bf
  2}(3) 033298
  \urlprefix\url{https://link.aps.org/doi/10.1103/PhysRevResearch.2.033298}

\bibitem{wlazlowski2023dissipation}
Wlaz\l{}owski G, Xhani K, Tylutki M, Proukakis N~P and Magierski P 2023 {\em
  Phys. Rev. Lett.\/} {\bf 130}(2) 023003
  \urlprefix\url{https://link.aps.org/doi/10.1103/PhysRevLett.130.023003}

\bibitem{klejdja2023decay}
Xhani K, Pace G~D, Scazza F and Roati G 2023  (\textit{Preprint}
  \eprint{arXiv:2306.11645})
  \urlprefix\url{https://doi.org/10.48550/arXiv.2306.11645}

\bibitem{ramanathan2011superflow}
Ramanathan A, Wright K~C, Muniz S~R, Zelan M, Hill W~T, Lobb C~J, Helmerson K,
  Phillips W~D and Campbell G~K 2011 {\em Physical Review Letters\/} {\bf
  106}(13) 130401
  \urlprefix\url{https://link.aps.org/doi/10.1103/PhysRevLett.106.130401}

\bibitem{eckel2014interferometric}
Eckel S, Jendrzejewski F, Kumar A, Lobb C~J and Campbell G~K 2014 {\em Physical
  Review X\/} {\bf 4} 031052
  \urlprefix\url{https://doi.org/10.1103/PhysRevX.4.031052}

\bibitem{ryu2013experimental}
Ryu C, Blackburn P, Blinova A and Boshier M 2013 {\em Physical Review
  Letters\/} {\bf 111} 205301
  \urlprefix\url{https://doi.org/10.1103/PhysRevLett.111.205301}

\bibitem{polo2022quantum}
Polo J, Naldesi P, Minguzzi A and Amico L 2021 {\em Quantum Science and
  Technology\/} {\bf 7} 015015
  \urlprefix\url{https://dx.doi.org/10.1088/2058-9565/ac39f6}

\bibitem{piazza2009vortex}
Piazza F, Collins L~A and Smerzi A 2009 {\em Physical Review A\/} {\bf 80}(2)
  021601 \urlprefix\url{https://link.aps.org/doi/10.1103/PhysRevA.80.021601}

\bibitem{qi2017magnetically}
Qi L, Hu Z, Valenzuela T, Zhang Y, Zhai Y, Quan W, Waltham N and Fang J 2017
  {\em Applied Physics Letters\/} {\bf 110} 153502
  \urlprefix\url{https://doi.org/10.1063/1.4980066}

\bibitem{wu2007demonstration}
Wu S, Su E and Prentiss M 2007 {\em Physical Review Letters\/} {\bf 99}
  \urlprefix\url{https://doi.org/10.1103/PhysRevLett.99.173201}

\bibitem{burke2009scalable}
Burke J~H~T and Sackett C~A 2009 {\em Physical Review A\/} {\bf 80} 061603
  \urlprefix\url{https://doi.org/10.1103/PhysRevA.80.061603}

\bibitem{burke2008confinement}
Burke J~H~T, Deissler B, Hughes K~J and Sackett C~A 2008 {\em Physical Review
  A\/} {\bf 78} \urlprefix\url{https://doi.org/10.1103/PhysRevA.78.023619}

\bibitem{deissler2008measurement}
Deissler B, Hughes K, Burke J and Sackett C 2008 {\em Physical Review A\/} {\bf
  77} 031604 \urlprefix\url{https://doi.org/10.1103/PhysRevA.77.031604}

\bibitem{arimondo2009atom}
Arimondo E, Ertmer W, Schleich W and Rasel E 2009 {\em {Atom Optics and Space
  Physics: Proceedings of the International School of Physics" Enrico Fermi",
  Course CLXVIII, Varenna on Lake Como, Villa Monastero, 3-13 July 2007}\/} vol
  168 (IOS Press)

\bibitem{bongs2019taking}
Bongs K, Holynski M, Vovrosh J, Bouyer P, Condon G, Rasel E, Schubert C,
  Schleich W~P and Roura A 2019 {\em Nature Reviews Physics\/} {\bf 1} 731--739
  \urlprefix\url{https://doi.org/10.1038/s42254-019-0117-4}

\bibitem{geiger2020high}
Geiger R, Landragin A, Merlet S and Pereira Dos~Santos F 2020 {\em AVS Quantum
  Science\/} {\bf 2} 024702 \urlprefix\url{https://doi.org/10.1116/5.0009093}

\bibitem{geiger2011detecting}
Geiger R, M\'{e}noret V, Stern G, Zahzam N, Cheinet P, Battelier B, Villing A,
  Moron F, Lours M, Bidel Y {\em et~al.\/} 2011 {\em Nature communications\/}
  {\bf 2} 1--7 \urlprefix\url{https://doi.org/10.1038/ncomms1479}

\bibitem{muller2008atom}
M\"{u}ller H, Chiow S~w, Long Q, Herrmann S and Chu S 2008 {\em Physical Review
  Letters\/} {\bf 100} 180405
  \urlprefix\url{https://doi.org/10.1103/PhysRevLett.100.180405}

\bibitem{stockton2011absolute}
Stockton J, Takase K and Kasevich M 2011 {\em Physical Review Letters\/} {\bf
  107} 133001 \urlprefix\url{https://doi.org/10.1103/PhysRevLett.107.133001}

\bibitem{sugarbaker2014atom}
Sugarbaker A 2014 {\em Atom interferometry in a 10 m fountain\/} (Stanford
  University) \urlprefix\url{https://purl.stanford.edu/kd753jv6128}

\bibitem{kovachy2015matter}
Kovachy T, Hogan J~M, Sugarbaker A, Dickerson S~M, Donnelly C~A, Overstreet C
  and Kasevich M~A 2015 {\em Physical Review Letters\/} {\bf 114}(14) 143004
  \urlprefix\url{https://doi.org/10.1103/PhysRevLett.114.143004}

\bibitem{muntinga2013interferometry}
Muntinga H, Ahlers H, Krutzik M, Wenzlawski A, Arnold S, Becker D, Bongs K,
  Dittus H, Duncker H, Gaaloul N, Gherasim C, Giese E, Grzeschik C, H\"ansch
  T~W, Hellmig O {\em et~al.\/} 2013 {\em Physical Review Letters\/} {\bf 110}
  093602 \urlprefix\url{https://doi.org/10.1103/PhysRevLett.110.093602}

\bibitem{helm2015sagnac}
Helm J~L, Cornish S~L and Gardiner S~A 2015 {\em Physical Review Letters\/}
  {\bf 114} 134101
  \urlprefix\url{https://doi.org/10.1103/PhysRevLett.114.134101}

\bibitem{helm2012bright}
Helm J~L, Billam T~P and Gardiner S~A 2012 {\em Physical Review A\/} {\bf 85}
  053621 \urlprefix\url{https://doi.org/10.1103/PhysRevA.85.053621}

\bibitem{mcdonald2014bright}
McDonald G~D, Kuhn C~C, Hardman K~S, Bennetts S, Everitt P~J, Altin P~A, Debs
  J~E, Close J~D and Robins N~P 2014 {\em Physical Review Letters\/} {\bf 113}
  013002 \urlprefix\url{https://doi.org/10.1103/PhysRevLett.113.013002}

\bibitem{polo2013soliton}
Polo J and Ahufinger V 2013 {\em Physical Review A\/} {\bf 88} 053628
  \urlprefix\url{https://doi.org/10.1103/PhysRevA.88.053628}

\bibitem{marchukov2019splitting}
Marchukov O~V, Malomed B~A, Yurovsky V~A, Olshanii M, Dunjko V and Hulet R~G
  2019 {\em Physical Review A\/} {\bf 99}(6) 063623
  \urlprefix\url{https://link.aps.org/doi/10.1103/PhysRevA.99.063623}

\bibitem{krzyzanowska2022matter}
Krzyzanowska K~A, Ferreras J, Ryu C, Samson E~C and Boshier M~G 2023 {\em
  Physical Review A\/} {\bf 108} ISSN 2469-9934
  \urlprefix\url{http://dx.doi.org/10.1103/PhysRevA.108.043305}

\bibitem{kim2022one}
Kim H, Krzyzanowska K, Henderson K~C, Ryu C, Timmermans E and Boshier M 2022
  {\em ArXiv e-prints\/} (\textit{Preprint} \eprint{2201.11888})
  \urlprefix\url{https://doi.org/10.48550/arXiv.2201.11888}

\bibitem{weiss2009creation}
Weiss C and Castin Y 2009 {\em Physical Review Letters\/} {\bf 102} 010403
  \urlprefix\url{https://doi.org/10.1103/PhysRevLett.102.010403}

\bibitem{streltsov2009scattering}
Streltsov A~I, Alon O~E and Cederbaum L~S 2009 {\em Physical Review A\/} {\bf
  80} 043616 \urlprefix\url{https://doi.org/10.1103/PhysRevA.80.043616}

\bibitem{haine2018quantum}
Haine S~A 2018 {\em New Journal of Physics\/} {\bf 20} 033009
  \urlprefix\url{https://iopscience.iop.org/article/10.1088/1367-2630/aab47f}

\bibitem{naldesi2022massive}
Naldesi P, Polo J, Drummond P~D, Dunjko V, Amico L, Minguzzi A and Olshanii M
  2023 {\em SciPost Physics\/} {\bf 15} 187
  \urlprefix\url{https://scipost.org/10.21468/SciPostPhys.15.5.187}

\bibitem{riedel2010atom}
Riedel M~F, B{\"o}hi P, Li Y, H{\"a}nsch T~W, Sinatra A and Treutlein P 2010
  {\em Nature\/} {\bf 464} 1170--1173
  \urlprefix\url{https://doi.org/10.1038/nature08988}

\bibitem{jo2007phase}
Jo G~B, Choi J~H, Christensen C~A, Pasquini T~A, Lee Y~R, Ketterle W and
  Pritchard D~E 2007 {\em Physical Review Letters\/} {\bf 98} 180401
  \urlprefix\url{https://doi.org/10.1103/PhysRevLett.98.180401}

\bibitem{schumm2005matter}
Schumm T, Hofferberth S, Andersson L~M, Wildermuth S, Groth S, Bar-Joseph I,
  Schmiedmayer J and Kr\"{u}ger P 2005 {\em Nature Physics\/} {\bf 1} 57--62
  \urlprefix\url{https://doi.org/10.1038/nphys125}

\bibitem{gunther2007atom}
G{\"u}nther A, Kraft S, Zimmermann C and Fort{\'a}gh J 2007 {\em Physical
  Review Letters\/} {\bf 98} 140403
  \urlprefix\url{https://doi.org/10.1103/PhysRevLett.98.140403}

\bibitem{bohi2009coherent}
B\"{o}hi P, Riedel M~F, Hoffrogge J, Reichel J, H\"{a}nsch T~W and Treutlein P
  2009 {\em Nature Physics\/} {\bf 5} 592--597
  \urlprefix\url{https://doi.org/10.1038/nphys1329}

\bibitem{akatsuka2017optically}
Akatsuka T, Takahashi T and Katori H 2017 {\em Applied Physics Express\/} {\bf
  10} 112501
  \urlprefix\url{https://iopscience.iop.org/article/10.7567/APEX.10.112501}

\bibitem{ryu2015integrated}
Ryu C and Boshier M~G 2015 {\em New Journal of Physics\/} {\bf 17} 092002
  \urlprefix\url{https://iopscience.iop.org/article/10.1088/1367-2630/17/9/092002}

\bibitem{moan2020quantum}
Moan E~R, Horne R~A, Arpornthip T, Luo Z, Fallon A~J, Berl S~J and Sackett C~A
  2020 {\em Physical Review Letters\/} {\bf 124}
  \urlprefix\url{https://doi.org/10.1103/PhysRevLett.124.120403}

\bibitem{navez2016matter}
Navez P, Pandey S, Mas H, Poulios K, Fernholz T and von Klitzing W 2016 {\em
  New Journal of Physics\/} {\bf 18} 075014
  \urlprefix\url{https://iopscience.iop.org/article/10.1088/1367-2630/18/7/075014}

\bibitem{stevenson2015sagnac}
Stevenson R, Hush M~R, Bishop T, Lesanovsky I and Fernholz T 2015 {\em Physical
  Review Letters\/} {\bf 115} 163001
  \urlprefix\url{https://doi.org/10.1103/PhysRevLett.115.163001}

\bibitem{halkyard2010rotational}
Halkyard P~L, Jones M~P~A and Gardiner S~A 2010 {\em Physical Review A\/} {\bf
  81} ISSN 1094-1622
  \urlprefix\url{http://dx.doi.org/10.1103/PhysRevA.81.061602}

\bibitem{japha2007using}
Japha Y, Arzouan O, Avishai Y and Folman R 2007 {\em Physical Review Letters\/}
  {\bf 99} 060402 \urlprefix\url{https://doi.org/10.1103/PhysRevLett.99.060402}

\bibitem{pelegri2018quantum}
Pelegr{\'{\i}} G, Mompart J and Ahufinger V 2018 {\em New Journal of Physics\/}
  {\bf 20} 103001
  \urlprefix\url{https://iopscience.iop.org/article/10.1088/1367-2630/aae107}

\bibitem{marti2015collective}
Marti G~E, Olf R and Stamper-Kurn D~M 2015 {\em Physical Review A\/} {\bf 91}
  013602 \urlprefix\url{https://doi.org/10.1103/PhysRevA.91.013602}

\bibitem{moukouri2021multipass}
Moukouri S, Japha Y, Keil M, David T, Groswasser D, Givon M and Folman R 2021
  {Multi-pass guided atomic Sagnac interferometer for high-performance rotation
  sensing} (\textit{Preprint} \eprint{2107.03446})
  \urlprefix\url{https://doi.org/10.48550/arXiv.2107.03446}

\bibitem{gauthier2019quantitative}
Gauthier G, Szigeti S~S, Reeves M~T, Baker M, Bell T~A, Rubinsztein-Dunlop H,
  Davis M~J and Neely T~W 2019 {\em Physical Review Letters\/} {\bf 123} 260402
  \urlprefix\url{https://doi.org/10.1103/PhysRevLett.123.260402}

\bibitem{lebrat2018band}
Lebrat M, Gri\ifmmode~\check{s}\else \v{s}\fi{}ins P, Husmann D, H\"ausler S,
  Corman L, Giamarchi T, Brantut J~P and Esslinger T 2018 {\em Physical Review
  X\/} {\bf 8} 011053 \urlprefix\url{https://doi.org/10.1103/PhysRevX.8.011053}

\bibitem{lebrat2019quantized}
Lebrat M, H\"ausler S, Fabritius P, Husmann D, Corman L and Esslinger T 2019
  {\em Physical Review Letters\/} {\bf 123} 193605
  \urlprefix\url{https://doi.org/10.1103/PhysRevLett.123.193605}

\bibitem{krinner2014observation}
Krinner S, Stadler D, Husmann D, Brantut J~P and Esslinger T 2014 {\em
  Nature\/} {\bf 517} 64--67
  \urlprefix\url{https://doi.org/10.1038/nature14049}

\bibitem{krinner2017two}
Krinner S, Esslinger T and Brantut J~P 2017 {\em Journal of Physics: Condensed
  Matter\/} {\bf 29} 343003
  \urlprefix\url{https://dx.doi.org/10.1088/1361-648X/aa74a1}

\bibitem{sommer2011universal}
Sommer A, Ku M, Roati G and Zwierlein M~W 2011 {\em Nature\/} {\bf 472}
  201--204 \urlprefix\url{https://doi.org/10.1038/nature09989}

\bibitem{brantut2012conduction}
Brantut J~P, Meineke J, Stadler D, Krinner S and Esslinger T 2012 {\em
  Science\/} {\bf 337} 1069--1071
  \urlprefix\url{https://www.science.org/doi/10.1126/science.1223175}

\bibitem{brantut2013thermoelectric}
Brantut J~P, Grenier C, Meineke J, Stadler D, Krinner S, Kollath C, Esslinger T
  and Georges A 2013 {\em Science\/} {\bf 342} 713--715
  \urlprefix\url{https://www.science.org/doi/10.1126/science.1242308}

\bibitem{husmann2015connecting}
Husmann D, Uchino S, Krinner S, Lebrat M, Giamarchi T, Esslinger T and Brantut
  J~P 2015 {\em Science\/} {\bf 350} 1498--1501 (\textit{Preprint}
  \eprint{https://www.science.org/doi/pdf/10.1126/science.aac9584})
  \urlprefix\url{https://www.science.org/doi/abs/10.1126/science.aac9584}

\bibitem{haug2019aharonov}
Haug T, Heimonen H, Dumke R, Kwek L~C and Amico L 2019 {\em Physical Review
  A\/} {\bf 100} 041601
  \urlprefix\url{https://doi.org/10.1103/PhysRevA.100.041601}

\bibitem{haug2019andreev}
Haug T, Dumke R, Kwek L~C and Amico L 2019 {\em Quantum Science and
  Technology\/} {\bf 4} 045001
  \urlprefix\url{https://iopscience.iop.org/article/10.1088/2058-9565/ab2e61}

\bibitem{delpace2021tunneling}
Del~Pace G, Kwon W~J, Zaccanti M, Roati G and Scazza F 2021 {\em Phys. Rev.
  Lett.\/} {\bf 126}(5) 055301
  \urlprefix\url{https://link.aps.org/doi/10.1103/PhysRevLett.126.055301}

\bibitem{stadler2012observing}
Stadler D, Krinner S, Meineke J, Brantut J~P and Esslinger T 2012 {\em
  Nature\/} {\bf 491} 736 \urlprefix\url{https://doi.org/10.1038/nature11613}

\bibitem{valtolina2015josephson}
Valtolina G, Burchianti A, Amico A, Neri E, Xhani K, Seman J~A, Trombettoni A,
  Smerzi A, Zaccanti M, Inguscio M and Roati G 2015 {\em Science\/} {\bf 350}
  1505--1508
  \urlprefix\url{https://www.science.org/doi/abs/10.1126/science.aac9725}

\bibitem{barontini2013controlling}
Barontini G, Labouvie R, Stubenrauch F, Vogler A, Guarrera V and Ott H 2013
  {\em Physical Review Letters\/} {\bf 110} 035302
  \urlprefix\url{https://doi.org/10.1103/PhysRevLett.110.035302}

\bibitem{labouvie2015negatice}
Labouvie R, Santra B, Heun S, Wimberger S and Ott H 2015 {\em Physical Review
  Letters\/} {\bf 115} 050601
  \urlprefix\url{https://doi.org/10.1103/PhysRevLett.115.050601}

\bibitem{pezze2018quantum}
Pezz\`e L, Smerzi A, Oberthaler M~K, Schmied R and Treutlein P 2018 {\em
  Reviews of Modern Physics\/} {\bf 90} 035005
  \urlprefix\url{https://doi.org/10.1103/RevModPhys.90.035005}

\bibitem{haroche2006exploring}
Haroche S and Raimond J~M 2006 {\em Exploring the quantum: atoms, cavities, and
  photons\/} (Oxford University Press)
  \urlprefix\url{https://doi.org/10.1093/acprof:oso/9780198509141.001.0001}

\bibitem{leBlanc2011dynamics}
LeBlanc L~J, Bardon A~B, McKeever J, Extavour M~H~T, Jervis D, Thywissen J~H,
  Piazza F and Smerzi A 2011 {\em Physical Review Letters\/} {\bf 106} 025302
  \urlprefix\url{https://doi.org/10.1103/PhysRevLett.106.025302}

\bibitem{spagnolli2017crossing}
Spagnolli G, Semeghini G, Masi L, Ferioli G, Trenkwalder A, Coop S, Landini M,
  Pezz\`e L, Modugno G, Inguscio M, Smerzi A and Fattori M 2017 {\em Physical
  Review Letters\/} {\bf 118} 230403
  \urlprefix\url{https://doi.org/10.1103/PhysRevLett.118.230403}

\bibitem{albiez2005direct}
Albiez M, Gati R, F\"olling J, Hunsmann S, Cristiani M and Oberthaler M~K 2005
  {\em Physical Review Letters\/} {\bf 95} 010402
  \urlprefix\url{https://doi.org/10.1103/PhysRevLett.95.010402}

\bibitem{levy2007the}
Levy S, Lahoud E, Shomroni I and Steinhauer J 2007 {\em Nature\/} {\bf 449}
  579--583 \urlprefix\url{https://doi.org/10.1038/nature06186}

\bibitem{pigneur2018relaxation}
Pigneur M, Berrada T, Bonneau M, Schumm T, Demler E and Schmiedmayer J 2018
  {\em Physical Review Letters\/} {\bf 120} 173601
  \urlprefix\url{https://doi.org/10.1103/PhysRevLett.120.173601}

\bibitem{gati2006noise}
Gati R, Hemmerling B, F\"{o}lling J, Albiez M and Oberthaler M~K 2006 {\em
  Physical Review Letters\/} {\bf 96} 130404
  \urlprefix\url{https://doi.org/10.1103/PhysRevLett.96.130404}

\bibitem{trenkwalder2016quantum}
Trenkwalder A, Spagnolli G, Semeghini G, Coop S, Landini M, Castilho P,
  Pezz\`{e} L, Modugno G, Inguscio M, Smerzi A and Fattori M 2016 {\em Nature
  Physics\/} {\bf 12} 826--829
  \urlprefix\url{https://doi.org/10.1038/nphys3743}

\bibitem{gaunt2012robust}
Gaunt A~L and Hadzibabic Z 2012 {\em Scientific Reports\/} {\bf 2} 721
  \urlprefix\url{https://doi.org/10.1038/srep00721}

\bibitem{mcgloin2003applications}
McGloin D, Spalding G~C, Melville H, Sibbett W and Dholakia K 2003 {\em Optics
  Express\/} {\bf 11} 158--166
  \urlprefix\url{https://doi.org/10.1364/OE.11.000158}

\bibitem{daley2008andreev}
Daley A, Zoller P and Trauzettel B 2008 {\em Physical Review Letters\/} {\bf
  100} 110404 \urlprefix\url{https://doi.org/10.1103/PhysRevLett.100.110404}

\bibitem{watabe2008reflection}
Watabe S and Kato Y 2008 {\em Physical Review A\/} {\bf 78} 063611
  \urlprefix\url{https://doi.org/10.1103/PhysRevA.78.063611}

\bibitem{zapata2009andreev}
Zapata I and Sols F 2009 {\em Physical Review Letters\/} {\bf 102} 180405
  \urlprefix\url{https://doi.org/10.1103/PhysRevLett.102.180405}

\bibitem{tokuno2008dynamics}
Tokuno A, Oshikawa M and Demler E 2008 {\em Physical Review Letters\/} {\bf
  100} 140402 \urlprefix\url{https://doi.org/10.1103/PhysRevLett.100.140402}

\bibitem{damanet2019controlling}
Damanet F, Mascarenhas E, Pekker D and Daley A~J 2019 {\em Physical Review
  Letters\/} {\bf 123} 180402

\bibitem{corman2019quantized}
Corman L, Fabritius P, H\"ausler S, Mohan J, Dogra L~H, Husmann D, Lebrat M and
  Esslinger T 2019 {\em Physical Review A\/} {\bf 100} 053605
  \urlprefix\url{https://doi.org/10.1103/PhysRevA.100.053605}

\bibitem{mullers2018coherent}
M\"{u}llers A, Santra B, Baals C, Jiang J, Benary J, Labouvie R, Zezyulin D~A,
  Konotop V~V and Ott H 2018 {\em Science Advances\/} {\bf 4} ISSN 2375-2548
  \urlprefix\url{http://dx.doi.org/10.1126/sciadv.aat6539}

\bibitem{filippone2016violation}
Filippone M, Hekking F and Minguzzi A 2016 {\em Physical Review A\/} {\bf 93}
  011602 \urlprefix\url{https://doi.org/10.1103/PhysRevA.93.011602}

\bibitem{husmann2018breakdown}
Husmann D, Lebrat M, H\"{a}usler S, Brantut J~P, Corman L and Esslinger T 2018
  {\em Proceedings of the National Academy of Sciences\/}
  \urlprefix\url{https://doi.org/10.1073/pnas.1803336115}

\bibitem{grenier2014peltier}
Grenier C, Georges A and Kollath C 2014 {\em Physical Review Letters\/} {\bf
  113} 200601 \urlprefix\url{https://doi.org/10.1103/PhysRevLett.113.200601}

\bibitem{sekera2016thermoelectricity}
Sekera T, Bruder C and Belzig W 2016 {\em Physical Review A\/} {\bf 94} 033618
  \urlprefix\url{https://doi.org/10.1103/PhysRevA.94.033618}

\bibitem{pershoguba2019thermopower}
Pershoguba S~S and Glazman L~I 2019 {\em Physical Review B\/} {\bf 99} 134514
  \urlprefix\url{https://doi.org/10.1103/PhysRevB.99.134514}

\bibitem{celi2014synthetic}
Celi A, Massignan P, Ruseckas J, Goldman N, Spielman I~B,
  Juzeli\ifmmode~\bar{u}\else \={u}\fi{}nas G and Lewenstein M 2014 {\em
  Physical Review Letters\/} {\bf 112} 043001
  \urlprefix\url{https://doi.org/10.1103/PhysRevLett.112.043001}

\bibitem{uchino2017anomalous}
Uchino S and Ueda M 2017 {\em Physical Review Letters\/} {\bf 118} 105303
  \urlprefix\url{https://doi.org/10.1103/PhysRevLett.118.105303}

\bibitem{Kanasz2016anomalous}
Kan\'asz-Nagy M, Glazman L, Esslinger T and Demler E~A 2016 {\em Physical
  Review Letters\/} {\bf 117} 255302
  \urlprefix\url{https://doi.org/10.1103/PhysRevLett.117.255302}

\bibitem{liu2017anomalous}
Liu B, Zhai H and Zhang S 2017 {\em Physical Review A\/} {\bf 95} 013623
  \urlprefix\url{https://doi.org/10.1103/PhysRevA.95.013623}

\bibitem{yao2018controlled}
Yao J, Liu B, Sun M and Zhai H 2018 {\em Physical Review A\/} {\bf 98} 041601
  \urlprefix\url{https://doi.org/10.1103/PhysRevA.98.041601}

\bibitem{gutman2012cold}
Gutman D~B, Gefen Y and Mirlin A~D 2012 {\em Physical Review B\/} {\bf 85}
  125102 \urlprefix\url{https://doi.org/10.1103/PhysRevB.85.125102}

\bibitem{simpson2014one}
Simpson D, Gangardt D, Lerner I and Kr\"{u}ger P 2014 {\em Physical review
  letters\/} {\bf 112} 100601
  \urlprefix\url{https://doi.org/10.1103/PhysRevLett.112.100601}

\bibitem{salerno2019quantized}
Salerno G, Price H~M, Lebrat M, H\"ausler S, Esslinger T, Corman L, Brantut J~P
  and Goldman N 2019 {\em Physical Review X\/} {\bf 9} 041001
  \urlprefix\url{https://doi.org/10.1103/PhysRevX.9.041001}

\bibitem{you2019atomtronics}
You J~S, Schmidt R, Ivanov D~A, Knap M and Demler E 2019 {\em Physical Review
  B\/} {\bf 99} 214505
  \urlprefix\url{https://doi.org/10.1103/PhysRevB.99.214505}

\bibitem{price2017synthetic}
Price H~M, Ozawa T and Goldman N 2017 {\em Physical Review A\/} {\bf 95} 023607
  \urlprefix\url{https://doi.org/10.1103/PhysRevA.95.023607}

\bibitem{hausler2017scanning}
H\"ausler S, Nakajima S, Lebrat M, Husmann D, Krinner S, Esslinger T and
  Brantut J~P 2017 {\em Physical Review Letters\/} {\bf 119} 030403
  \urlprefix\url{https://doi.org/10.1103/PhysRevLett.119.030403}

\bibitem{white2019observation}
White D~H, Haase T~A, Brown D~J, Hoogerland M~D, Najafabadi M~S, Helm J~L, Gies
  C, Schumayer D and Hutchinson D~A~W 2020 {\em Nature Communications\/} {\bf
  11} ISSN 2041-1723
  \urlprefix\url{http://dx.doi.org/10.1038/s41467-020-18652-w}

\bibitem{bruderer2012mesoscopic}
Bruderer M and Belzig W 2012 {\em Physical Review A\/} {\bf 85} 013623
  \urlprefix\url{https://doi.org/10.1103/PhysRevA.85.013623}

\bibitem{chien2014landauer}
Chien C~C, Di~Ventra M and Zwolak M 2014 {\em Physical Review A\/} {\bf 90}
  023624 \urlprefix\url{https://doi.org/10.1103/PhysRevA.90.023624}

\bibitem{gallego2014nonequilibrium}
Gallego-Marcos F, Platero G, Nietner C, Schaller G and Brandes T 2014 {\em
  Physical Review A\/} {\bf 90} 033614
  \urlprefix\url{https://doi.org/10.1103/PhysRevA.90.033614}

\bibitem{ivanov2013bosonic}
Ivanov A, Kordas G, Komnik A and Wimberger S 2013 {\em The European Physical
  Journal B\/} {\bf 86} 345
  \urlprefix\url{https://doi.org/10.1140/epjb/e2013-40417-4}

\bibitem{kolovsky2017microscopic}
Kolovsky A~R 2017 {\em Physical Review A\/} {\bf 96} 011601
  \urlprefix\url{https://doi.org/10.1103/PhysRevA.96.011601}

\bibitem{nietner2014transport}
Nietner C, Schaller G and Brandes T 2014 {\em Physical Review A\/} {\bf 89}
  013605 \urlprefix\url{https://doi.org/10.1103/PhysRevA.89.013605}

\bibitem{kolovsky2018landauer}
Kolovsky A~R, Denis Z and Wimberger S 2018 {\em Physical Review A\/} {\bf 98}
  043623 \urlprefix\url{https://doi.org/10.1103/PhysRevA.98.043623}

\bibitem{papoular2016quantized}
Papoular D~J, Pitaevskii L~P and Stringari S 2016 {\em Physical Review A\/}
  {\bf 94} 023622 \urlprefix\url{https://doi.org/10.1103/PhysRevA.94.023622}

\bibitem{cuevas2017molecular}
Cuevas J~C and Scheer E 2017 {\em {Molecular Electronics}\/} vol Volume 15
  (World Scientific) \urlprefix\url{https://doi.org/10.1142/10598}

\bibitem{krinner2013superfluidity}
Krinner S, Stadler D, Meineke J, Brantut J~P and Esslinger T 2013 {\em Physical
  Review Letters\/} {\bf 110} 100601
  \urlprefix\url{https://doi.org/10.1103/PhysRevLett.110.100601}

\bibitem{krinner2015observation}
Krinner S, Stadler D, Meineke J, Brantut J~P and Esslinger T 2015 {\em Physical
  Review Letters\/} {\bf 115} 045302
  \urlprefix\url{https://doi.org/10.1103/PhysRevLett.115.045302}

\bibitem{krinner2015observationa}
Krinner S, Stadler D, Husmann D, Brantut J~P and Esslinger T 2015 {\em
  Nature\/} {\bf 517} 64--67
  \urlprefix\url{https://doi.org/10.1038/nature14049}

\bibitem{pyykkonen2021flat}
Pyykk\"onen V~A~J, Peotta S, Fabritius P, Mohan J, Esslinger T and T\"orm\"a P
  2021 {\em Physical Review B\/} {\bf 103} 144519
  \urlprefix\url{https://doi.org/10.1103/PhysRevB.103.144519}

\bibitem{kwon2020strongly}
Kwon W~J, Pace G~D, Panza R, Inguscio M, Zwerger W, Zaccanti M, Scazza F and
  Roati G 2020 {\em Science\/} {\bf 369} 84--88
  \urlprefix\url{https://www.science.org/doi/abs/10.1126/science.aaz2463}

\bibitem{zaccanti2019critical}
Zaccanti M and Zwerger W 2019 {\em Physical Review A\/} {\bf 100} 063601
  \urlprefix\url{https://doi.org/10.1103/PhysRevA.100.063601}

\bibitem{uchino2020bosonic}
Uchino S and Brantut J~P 2020 {\em Physical Review Research\/} {\bf 2} 023284
  \urlprefix\url{https://doi.org/10.1103/PhysRevResearch.2.023284}

\bibitem{uchino2020role}
Uchino S 2020 {\em Physical Review Research\/} {\bf 2} 023340
  \urlprefix\url{https://doi.org/10.1103/PhysRevResearch.2.023340}

\bibitem{luick2020ideal}
Luick N, Sobirey L, Bohlen M, Singh V~P, Mathey L, Lompe T and Moritz H 2020
  {\em Science\/} {\bf 369} 89
  \urlprefix\url{https://www.science.org/doi/abs/10.1126/science.aaz2342}

\bibitem{tononi2020dephasing}
Tononi A, Toigo F, Wimberger S, Cappellaro A and Salasnich L 2020 {\em New
  Journal of Physics\/} {\bf 22} 073020
  \urlprefix\url{https://iopscience.iop.org/article/10.1088/1367-2630/ab965d}

\bibitem{binanti2021dissipation}
Binanti F, Furutani K and Salasnich L 2021 {\em Physical Review A\/} {\bf 103}
  ISSN 2469-9934 \urlprefix\url{http://dx.doi.org/10.1103/PhysRevA.103.063309}

\bibitem{schweigler2021decay}
Schweigler T, Gluza M, Tajik M, Sotiriadis S, Cataldini F, Ji S~C, M{\o}ller
  F~S, Sabino J, Rauer B, Eisert J and Schmiedmayer J 2021 {\em Nature
  Physics\/} {\bf 17} 559--563
  \urlprefix\url{https://doi.org/10.1038/s41567-020-01139-2}

\bibitem{langen2015ultracold}
Langen T, Geiger R and Schmiedmayer J 2015 {\em Annual Review of Condensed
  Matter Physics\/} {\bf 6} 201--217
  \urlprefix\url{https://doi.org/10.1146/annurev-conmatphys-031214-014548}

\bibitem{betz2011two}
Betz T, Manz S, B\"ucker R, Berrada T, Koller C, Kazakov G, Mazets I~E,
  Stimming H~P, Perrin A, Schumm T and Schmiedmayer J 2011 {\em Physical Review
  Letters\/} {\bf 106} 020407
  \urlprefix\url{https://doi.org/10.1103/PhysRevLett.106.020407}

\bibitem{gring2012relaxation}
Gring M, Kuhnert M, Langen T, Kitagawa T, Rauer B, Schreitl M, Mazets I, Smith
  D~A, Demler E and Schmiedmayer J 2012 {\em Science\/} {\bf 337} 1318--1322
  \urlprefix\url{https://www.science.org/doi/10.1126/science.1224953}

\bibitem{hofferberth2007nonequilibrium}
Hofferberth S, Lesanovsky I, Fischer B, Schumm T and Schmiedmayer J 2007 {\em
  Nature\/} {\bf 449} 324--327
  \urlprefix\url{https://doi.org/10.1038/nature06149}

\bibitem{hofferberth2008probing}
Hofferberth S, Lesanovsky I, Schumm T, Imambekov A, Gritsev V, Demler E and
  Schmiedmayer J 2008 {\em Nature Physics\/} {\bf 4} 489--495
  \urlprefix\url{https://doi.org/10.1038/nphys941}

\bibitem{giovanazzi2008effective}
Giovanazzi S, Esteve J and Oberthaler M~K 2008 {\em New Journal of Physics\/}
  {\bf 10} 045009 \urlprefix\url{https://doi.org/10.1088/1367-2630/10/4/045009}

\bibitem{fukuhara2007degenerate}
Fukuhara T, Takasu Y, Kumakura M and Takahashi Y 2007 {\em Physical Review
  Letters\/} {\bf 98}(3) 030401
  \urlprefix\url{https://link.aps.org/doi/10.1103/PhysRevLett.98.030401}

\bibitem{fukuhara2007bose}
Fukuhara T, Sugawa S and Takahashi Y 2007 {\em Physical Review A\/} {\bf 76}(5)
  051604 \urlprefix\url{https://link.aps.org/doi/10.1103/PhysRevA.76.051604}

\bibitem{fukuhara2009all}
Fukuhara T, Sugawa S, Takasu Y and Takahashi Y 2009 {\em Physical Review A\/}
  {\bf 79}(2) 021601
  \urlprefix\url{https://link.aps.org/doi/10.1103/PhysRevA.79.021601}

\bibitem{cazalilla2009ultracold}
Cazalilla M~A, Ho A and Ueda M 2009 {\em New Journal of Physics\/} {\bf 11}
  103033

\bibitem{ho1998spinor}
Ho T~L 1998 {\em Physical Review Letters\/} {\bf 81}(4) 742--745
  \urlprefix\url{https://link.aps.org/doi/10.1103/PhysRevLett.81.742}

\bibitem{ohmi1998bose}
Ohmi T and Machida K 1998 {\em Journal of the Physical Society of Japan\/} {\bf
  67} 1822–1825 ISSN 1347-4073
  \urlprefix\url{http://dx.doi.org/10.1143/JPSJ.67.1822}

\bibitem{yip1999zero}
Yip S~K and Ho T~L 1999 {\em Physical Review A\/} {\bf 59}(6) 4653--4656
  \urlprefix\url{https://link.aps.org/doi/10.1103/PhysRevA.59.4653}

\bibitem{stamper2013spinor}
Stamper-Kurn D~M and Ueda M 2013 {\em Reviews of Modern Physics\/} {\bf 85}(3)
  1191--1244
  \urlprefix\url{https://link.aps.org/doi/10.1103/RevModPhys.85.1191}

\bibitem{cazalilla2014ultracold}
Cazalilla M~A and Rey A~M 2014 {\em Reports on Progress in Physics\/} {\bf 77}
  124401 \urlprefix\url{https://doi.org/10.1088/0034-4885/77/12/124401}

\bibitem{mistakidis2023few}
Mistakidis S, Volosniev A, Barfknecht R, Fogarty T, Busch T, Foerster A,
  Schmelcher P and Zinner N 2023 {\em Physics Reports\/} {\bf 1042} 1--108 ISSN
  0370-1573 {Few-body Bose gases in low dimensions—A laboratory for quantum
  dynamics}
  \urlprefix\url{https://www.sciencedirect.com/science/article/pii/S0370157323003162}

\bibitem{gorshkov2010two}
Gorshkov A~V, Hermele M, Gurarie V, Xu C, Julienne P~S, Ye J, Zoller P, Demler
  E, Lukin M~D and Rey A~M 2010 {\em Nature Physics\/} {\bf 6} 289–295 ISSN
  1745-2481 \urlprefix\url{http://dx.doi.org/10.1038/nphys1535}

\bibitem{capponi2016phases}
Capponi S, Lecheminant P and Totsuka K 2016 {\em Annals of Physics\/} {\bf 367}
  50--95 \urlprefix\url{https://doi.org/10.1016/j.aop.2016.01.011}

\bibitem{scazza2014observation}
Scazza F, Hofrichter C, H\"{o}fer M, De~Groot P~C, Bloch I and F\"{o}lling S
  2014 {\em Nature Physics\/} {\bf 10} 779–784 ISSN 1745-2481
  \urlprefix\url{http://dx.doi.org/10.1038/nphys3061}

\bibitem{hofrichter2016direct}
Hofrichter C, Riegger L, Scazza F, H\"ofer M, Fernandes D~R, Bloch I and
  F\"olling S 2016 {\em Physical Review X\/} {\bf 6}(2) 021030
  \urlprefix\url{https://link.aps.org/doi/10.1103/PhysRevX.6.021030}

\bibitem{kolkowitz2016spin}
Kolkowitz S, Bromley S~L, Bothwell T, Wall M~L, Marti G~E, Koller A~P, Zhang X,
  Rey A~M and Ye J 2016 {\em Nature\/} {\bf 542} 66–70 ISSN 1476-4687
  \urlprefix\url{http://dx.doi.org/10.1038/nature20811}

\bibitem{taie2022observation}
Taie S, Ibarra-García-Padilla E, Nishizawa N, Takasu Y, Kuno Y, Wei H~T,
  Scalettar R~T, Hazzard K~R~A and Takahashi Y 2022 {\em Nature Physics\/} {\bf
  18} 1356–1361 ISSN 1745-2481
  \urlprefix\url{http://dx.doi.org/10.1038/s41567-022-01725-6}

\bibitem{takahashi2022quantum}
TAKAHASHI Y 2022 {\em Proceedings of the Japan Academy, Series B\/} {\bf 98}
  141–160 ISSN 1349-2896
  \urlprefix\url{http://dx.doi.org/10.2183/pjab.98.010}

\bibitem{ludlow2015optical}
Ludlow A~D, Boyd M~M, Ye J, Peik E and Schmidt P~O 2015 {\em Reviews of Modern
  Physics\/} {\bf 87}(2) 637--701
  \urlprefix\url{https://link.aps.org/doi/10.1103/RevModPhys.87.637}

\bibitem{marti2018imaging}
Marti G~E, Hutson R~B, Goban A, Campbell S~L, Poli N and Ye J 2018 {\em
  Physical Review Letters\/} {\bf 120}(10) 103201
  \urlprefix\url{https://link.aps.org/doi/10.1103/PhysRevLett.120.103201}

\bibitem{imry2002introduction}
Imry Y 2002 {\em Introduction to mesoscopic physics\/} (Oxford University
  Press)

\bibitem{giorgini2008theory}
Giorgini S, Pitaevskii L~P and Stringari S 2008 {\em Rev. Mod. Phys.\/} {\bf
  80}(4) 1215--1274
  \urlprefix\url{https://link.aps.org/doi/10.1103/RevModPhys.80.1215}

\bibitem{guan2013fermi}
Guan X~W, Batchelor M~T and Lee C 2013 {\em Reviews of Modern Physics\/} {\bf
  85} 1633 \urlprefix\url{https://doi.org/10.1103/RevModPhys.85.1633}

\bibitem{chetcuti2023probe}
Chetcuti W~J, Polo J, Osterloh A, Castorina P and Amico L 2023 {\em
  Communications Physics\/} {\bf 6}
  \urlprefix\url{https://doi.org/10.1038/s42005-023-01256-3}

\bibitem{capponi2008molecular}
Capponi S, Roux G, Lecheminant P, Azaria P, Boulat E and White S~R 2008 {\em
  Physical Review A\/} {\bf 77}(1) 013624
  \urlprefix\url{https://link.aps.org/doi/10.1103/PhysRevA.77.013624}

\bibitem{beattie2013persistent}
Beattie S, Moulder S, Fletcher R~J and Hadzibabic Z 2013 {\em Physical Review
  Letters\/} {\bf 110} ISSN 1079-7114
  \urlprefix\url{http://dx.doi.org/10.1103/PhysRevLett.110.025301}

\bibitem{anoshkin2013persistent}
Anoshkin K, Wu Z and Zaremba E 2013 {\em Physical Review A\/} {\bf 88}(1)
  013609 \urlprefix\url{https://link.aps.org/doi/10.1103/PhysRevA.88.013609}

\bibitem{white2017odd}
White A~C, Zhang Y and Busch T 2017 {\em Physical Review A\/} {\bf 95}(4)
  041604 \urlprefix\url{https://link.aps.org/doi/10.1103/PhysRevA.95.041604}

\bibitem{spehner2021persistent}
Spehner D, Morales-Molina L and Reyes S~A 2021 {\em New Journal of Physics\/}
  \urlprefix\url{https://doi.org/10.1088/1367-2630/abeebb}

\bibitem{richaud2021interaction}
Richaud A, Ferraretto M and Capone M 2021 {\em Physical Review B\/} {\bf 103}
  205132 \urlprefix\url{https://doi.org/10.1103/PhysRevB.103.205132}

\bibitem{chetcuti2022persistent}
Chetcuti W~J, Haug T, Kwek L~C and Amico L 2022 {\em SciPost Physics\/} {\bf
  12} 033 \urlprefix\url{https://scipost.org/10.21468/SciPostPhys.12.1.033}

\bibitem{consiglio2022variational}
Consiglio M, Chetcuti W~J, Bravo-Prieto C, Ramos-Calderer S, Minguzzi A,
  Latorre J~I, Amico L and Apollaro T~J~G 2022 {\em Journal of Physics A:
  Mathematical and Theoretical\/} {\bf 55} 265301
  \urlprefix\url{https://dx.doi.org/10.1088/1751-8121/ac7016}

\bibitem{richaud2022mimicking}
Richaud A, Ferraretto M and Capone M 2022 {\em Condensed Matter\/} {\bf 7} 18
  ISSN 2410-3896 \urlprefix\url{http://dx.doi.org/10.3390/condmat7010018}

\bibitem{chetcuti2023interference}
Chetcuti W~J, Osterloh A, Amico L and Polo J 2023 {\em SciPost Physics\/} {\bf
  15} 181 \urlprefix\url{https://scipost.org/10.21468/SciPostPhys.15.4.181}

\bibitem{osterloh2023exact}
Osterloh A, Polo J, Chetcuti W~J and Amico L 2023 {\em SciPost Physics\/} {\bf
  15} 006 \urlprefix\url{https://scipost.org/10.21468/SciPostPhys.15.1.006}

\bibitem{chetcuti2023persistent}
Chetcuti W~J 2023 {Persistent Currents in Atomtronic Circuits of SU($N$)
  Fermions} (\textit{Preprint} \eprint{2311.03072})

\bibitem{pecci2023persistent}
Pecci G, Aupetit-Diallo G, Albert M, Vignolo P and Minguzzi A 2023 {\em Comptes
  Rendus. Physique\/} {\bf 24} 1--13
  \urlprefix\url{https://doi.org/10.5802/crphys.157}

\bibitem{polo2020quantum}
Polo J, Naldesi P, Minguzzi A and Amico L 2021 {\em Quantum Science and
  Technology\/} {\bf 7} 015015
  \urlprefix\url{https://doi.org/10.1088/2058-9565/ac39f6}

\bibitem{kane1992transmission}
Kane C~L and Fisher M~P~A 1992 {\em Physical Review B\/} {\bf 46}(23)
  15233--15262
  \urlprefix\url{https://link.aps.org/doi/10.1103/PhysRevB.46.15233}

\bibitem{saleur1998lectures}
Saleur H 1998 {Lectures on Non Perturbative Field Theory and Quantum Impurity
  Problems} (\textit{Preprint} \eprint{cond-mat/9812110})

\bibitem{rylands2016quantum}
Rylands C and Andrei N 2016 {\em Physical Review B\/} {\bf 94}(11) 115142
  \urlprefix\url{https://link.aps.org/doi/10.1103/PhysRevB.94.115142}

\bibitem{cominotti2015scaling}
Cominotti M, Rizzi M, Rossini D, Aghamalyan D, Amico L, Kwek L~C, Hekking F and
  Minguzzi A 2015 {\em The European Physical Journal Special Topics\/} {\bf
  224} 519--524 \urlprefix\url{https://doi.org/10.1140/epjst/e2015-02381-3}

\bibitem{kiehn2022implementation}
Kiehn H, Singh V~P and Mathey L 2022 {\em Phys. Rev. Res.\/} {\bf 4}(3) 033024
  \urlprefix\url{https://link.aps.org/doi/10.1103/PhysRevResearch.4.033024}

\bibitem{aghamalyan2015atomtronics}
Aghamalyan D 2015 {\em {Atomtronics: Quantum technology with cold atoms in ring
  shaped optical lattices}\/} Ph.D. thesis National University of Singapore
  \urlprefix\url{https://quantumlah.org/media/thesis/CQT_151210_DavitAghamalyan.pdf}

\bibitem{aghamalyanatomtronic2016}
Aghamalyan D, Nguyen N~T, Auksztol F, Gan K~S, Valado M~M, Condylis P~C, Kwek
  L~C, Dumke R and Amico L 2016 {\em New Journal of Physics\/} {\bf 18} 075013
  \urlprefix\url{https://doi.org/10.1088/1367-2630/18/7/075013}

\bibitem{singh2023shapiro}
Singh V~P, Polo J, Mathey L and Amico L 2023 Shapiro steps in driven atomic
  josephson junctions (\textit{Preprint} \eprint{arXiv:2307.08743})

\bibitem{shapiro1963josephson}
Shapiro S 1963 {\em Phys. Rev. Lett.\/} {\bf 11}(2) 80--82
  \urlprefix\url{https://link.aps.org/doi/10.1103/PhysRevLett.11.80}

\bibitem{grimes1968millimeter}
Grimes C~C and Shapiro S 1968 {\em Phys. Rev.\/} {\bf 169}(2) 397--406
  \urlprefix\url{https://link.aps.org/doi/10.1103/PhysRev.169.397}

\bibitem{hamilton1995ieee}
Hamilton C, Burroughs C and Kautz R 1995 {\em IEEE Transactions on
  Instrumentation and Measurement\/} {\bf 44} 223--225
  \urlprefix\url{https://ieeexplore.ieee.org/document/377816}

\bibitem{burroughs1999ieee}
Burroughs C, Benz S, Harvey T and Hamilton C 1999 {\em IEEE Transactions on
  Applied Superconductivity\/} {\bf 9} 4145--4149
  \urlprefix\url{https://ieeexplore.ieee.org/document/783938}

\bibitem{burroughs2011ieee}
Burroughs C~J, Dresselhaus P~D, Rufenacht A, Olaya D, Elsbury M~M, Tang Y~H and
  Benz S~P 2011 {\em IEEE Transactions on Instrumentation and Measurement\/}
  {\bf 60} 2482--2488
  \urlprefix\url{https://ieeexplore.ieee.org/document/5686932}

\bibitem{saffman2010quantum}
Saffman M, Walker T~G and M{\o}lmer K 2010 {\em Reviews of Modern Physics\/}
  {\bf 82} 2313--2363
  \urlprefix\url{https://doi.org/10.1103/revmodphys.82.2313}

\bibitem{adams2019rydberg}
Adams C~S, Pritchard J~D and Shaffer J~P 2019 {\em Journal of Physics B:
  Atomic, Molecular and Optical Physics\/} {\bf 53} 012002
  \urlprefix\url{https://doi.org/10.1088/1361-6455/ab52ef}

\bibitem{browaeys2020many}
Browaeys A and Lahaye T 2020 {\em Nature Physics\/} {\bf 16} 132--142
  \urlprefix\url{https://doi.org/10.1038/s41567-019-0733-z}

\bibitem{wu2021concise}
Wu X, Liang X, Tian Y, Yang F, Chen C, Liu Y~C, Tey M~K and You L 2021 {\em
  Chinese Physics B\/} {\bf 30} 020305
  \urlprefix\url{https://doi.org/10.1088/1674-1056/abd76f}

\bibitem{morgado2021quantum}
Morgado M and Whitlock S 2021 {\em AVS Quantum Science\/} {\bf 3} ISSN
  2639-0213 \urlprefix\url{http://dx.doi.org/10.1116/5.0036562}

\bibitem{bernien2017probing}
Bernien H, Schwartz S, Keesling A, Levine H, Omran A, Pichler H, Choi S, Zibrov
  A~S, Endres M, Greiner M, Vuleti{\'{c}} V and Lukin M~D 2017 {\em Nature\/}
  {\bf 551} 579--584 \urlprefix\url{https://doi.org/10.1038/nature24622}

\bibitem{pohl2010dynamical}
Pohl T, Demler E and Lukin M~D 2010 {\em Phys. Rev. Lett.\/} {\bf 104}(4)
  043002
  \urlprefix\url{https://link.aps.org/doi/10.1103/PhysRevLett.104.043002}

\bibitem{barredo2015coherent}
Barredo D, Labuhn H, Ravets S, Lahaye T, Browaeys A and Adams C~S 2015 {\em
  Physical Review Letters\/} {\bf 114}
  \urlprefix\url{https://doi.org/10.1103/physrevlett.114.113002}

\bibitem{endres2016atom}
Endres M, Bernien H, Keesling A, Levine H, Anschuetz E~R, Krajenbrink A, Senko
  C, Vuletic V, Greiner M and Lukin M~D 2016 {\em Science\/} {\bf 354}
  1024--1027 \urlprefix\url{https://doi.org/10.1126/science.aah3752}

\bibitem{barredo2018synthetic}
Barredo D, Lienhard V, De~Leseleuc S, Lahaye T and Browaeys A 2018 {\em
  Nature\/} {\bf 561} 79--82
  \urlprefix\url{https://doi.org/10.1038/s41586-018-0450-2}

\bibitem{barredo2016atom}
Barredo D, de~L{\'{e}}s{\'{e}}leuc S, Lienhard V, Lahaye T and Browaeys A 2016
  {\em Science\/} {\bf 354} 1021--1023
  \urlprefix\url{https://doi.org/10.1126/science.aah3778}

\bibitem{schymik2020enhanced}
Schymik K~N, Lienhard V, Barredo D, Scholl P, Williams H, Browaeys A and Lahaye
  T 2020 {\em Physical Review A\/} {\bf 102}
  \urlprefix\url{https://doi.org/10.1103/physreva.102.063107}

\bibitem{lukin2001dipole}
Lukin M~D, Fleischhauer M, Cote R, Duan L~M, Jaksch D, Cirac J~I and Zoller P
  2001 {\em Phys. Rev. Lett.\/} {\bf 87}(3) 037901
  \urlprefix\url{https://link.aps.org/doi/10.1103/PhysRevLett.87.037901}

\bibitem{urban2009observation}
Urban E, Johnson T~A, Henage T, Isenhower L, Yavuz D~D, Walker T~G and Saffman
  M 2009 {\em Nature Physics\/} {\bf 5} 110–114 ISSN 1745-2481
  \urlprefix\url{http://dx.doi.org/10.1038/nphys1178}

\bibitem{valado2016experimental}
Valado M, Simonelli C, Hoogerland M, Lesanovsky I, Garrahan J~P, Arimondo E,
  Ciampini D and Morsch O 2016 {\em Physical Review A\/} {\bf 93} 040701
  \urlprefix\url{https://doi.org/10.1103/PhysRevA.93.040701}

\bibitem{morsch2018many}
Morsch O and Lesanovsky I 2018 {\em La Rivista del Nuovo Cimento\/} {\bf 41}
  383–414 \urlprefix\url{https://doi.org/10.1393/ncr/i2018-10149-7}

\bibitem{lesanovsky2014out}
Lesanovsky I and Garrahan J~P 2014 {\em Phys. Rev. A\/} {\bf 90}(1) 011603
  \urlprefix\url{https://link.aps.org/doi/10.1103/PhysRevA.90.011603}

\bibitem{bose2003quantum}
Bose S 2003 {\em Physical Review Letters\/} {\bf 91}
  \urlprefix\url{https://doi.org/10.1103/physrevlett.91.207901}

\bibitem{christandl2004perfect}
Christandl M, Datta N, Ekert A and Landahl A~J 2004 {\em Physical Review
  Letters\/} {\bf 92}
  \urlprefix\url{https://doi.org/10.1103/physrevlett.92.187902}

\bibitem{paganelli2013routing}
Paganelli S, Lorenzo S, Apollaro T~J~G, Plastina F and Giorgi G~L 2013 {\em
  Physical Review A\/} {\bf 87}
  \urlprefix\url{https://doi.org/10.1103/physreva.87.062309}

\bibitem{apollaro2015many}
Apollaro T~J~G, Lorenzo S, Sindona A, Paganelli S, Giorgi G~L and Plastina F
  2015 {\em Physica Scripta\/} {\bf 2015} 014036
  \urlprefix\url{https://dx.doi.org/10.1088/0031-8949/2015/T165/014036}

\bibitem{chetcuti2020perturbative}
Chetcuti W~J, Sanavio C, Lorenzo S and Apollaro T~J~G 2020 {\em New Journal of
  Physics\/} {\bf 22} 033030
  \urlprefix\url{https://doi.org/10.1088/1367-2630/ab7a33}

\bibitem{apollaro2020two}
Apollaro T~J~G and Chetcuti W~J 2020 {\em Entropy\/} {\bf 23} 51
  \urlprefix\url{https://doi.org/10.3390/e23010051}

\bibitem{apollaro2022quantum}
Apollaro T~J~G, Lorenzo S, Plastina F, Consiglio M and {\.{Z}}yczkowski K 2022
  {\em New Journal of Physics\/} {\bf 24} 083025
  \urlprefix\url{https://doi.org/10.1088/1367-2630/ac86e7}

\bibitem{palaiodimopulos2023chiral}
Palaiodimopoulos N~E, Ohler S, Fleischhauer M and Petrosyan D 2023 A chiral
  quantum router with rydberg atoms
  \urlprefix\url{https://arxiv.org/abs/2310.10390}

\bibitem{li2019shao}
Li D~X and Shao X~Q 2019 {\em Physical Review A\/} {\bf 99}
  \urlprefix\url{https://doi.org/10.1103/physreva.99.032348}

\bibitem{moffitt2008recent}
Moffitt J~R, Chemla Y~R, Smith S~B and Bustamante C 2008 {\em Annual Review of
  Biochemistry\/} {\bf 77} 205–228 ISSN 1545-4509
  \urlprefix\url{http://dx.doi.org/10.1146/annurev.biochem.77.043007.090225}

\bibitem{perciavalle2023coherent}
Perciavalle F, Morsch O, Rossini D and Amico L 2023 Coherent excitation
  transport through ring-shaped networks (\textit{Preprint}
  \eprint{2310.17967})

\bibitem{wu2022manipulating}
Wu X, Yang F, Yang S, M{\o}lmer K, Pohl T, Tey M~K and You L 2022 {\em Physical
  Review Research\/} {\bf 4}
  \urlprefix\url{https://doi.org/10.1103/physrevresearch.4.l032046}

\bibitem{perciavalle2023controlled}
Perciavalle F, Rossini D, Haug T, Morsch O and Amico L 2023 {\em Physical
  Review A\/} {\bf 108}(2) 023305
  \urlprefix\url{https://link.aps.org/doi/10.1103/PhysRevA.108.023305}

\bibitem{lienhard2020realization}
Lienhard V, Scholl P, Weber S, Barredo D, de~L{\'{e}}s{\'{e}}leuc S, Bai R,
  Lang N, Fleischhauer M, B\"{u}chler H~P, Lahaye T and Browaeys A 2020 {\em
  Physical Review X\/} {\bf 10}
  \urlprefix\url{https://doi.org/10.1103/physrevx.10.021031}

\bibitem{kitson2023rydberg}
Kitson P, Haug T, Magna A~L, Morsch O and Amico L 2023 Rydberg atomtronic
  devices (\textit{Preprint} \eprint{2310.18242})

\bibitem{begoc2023controlled}
Bégoc B, Cichelli G, Singh S~P, Perciavalle F, Rossini D, Amico L and Morsch O
  2023 {Controlled dissipation for Rydberg atom experiments} (\textit{Preprint}
  \eprint{2310.20687})

\bibitem{marcuzzi2016absorbing}
Marcuzzi M, Buchhold M, Diehl S and Lesanovsky I 2016 {\em Physical Review
  Letters\/} {\bf 116}
  \urlprefix\url{https://doi.org/10.1103/physrevlett.116.245701}

\bibitem{wintermantel2021epidemic}
Wintermantel T~M, Buchhold M, Shevate S, Morgado M, Wang Y, Lochead G, Diehl S
  and Whitlock S 2021 {\em Nature Communications\/} {\bf 12} ISSN 2041-1723
  \urlprefix\url{http://dx.doi.org/10.1038/s41467-020-20333-7}

\bibitem{zohar2016quantum}
Zohar E, Cirac J~I and Reznik B 2015 {\em Reports on Progress in Physics\/}
  {\bf 79} 014401
  \urlprefix\url{https://dx.doi.org/10.1088/0034-4885/79/1/014401}

\bibitem{mil2020scalable}
Mil A, Zache T~V, Hegde A, Xia A, Bhatt R~P, Oberthaler M~K, Hauke P, Berges J
  and Jendrzejewski F 2020 {\em Science\/} {\bf 367} 1128--1130
  \urlprefix\url{https://www.science.org/doi/10.1126/science.aaz5312}

\bibitem{yang2020observation}
Yang B, Sun H, Ott R, Wang H~Y, Zache T~V, Halimeh J~C, Yuan Z~S, Hauke P and
  Pan J~W 2020 {\em Nature\/} {\bf 587} 392--396
  \urlprefix\url{https://doi.org/10.1038/s41586-020-2910-8}

\bibitem{zhao2022thermalization}
Zhou Z~Y, Su G~X, Halimeh J~C, Ott R, Sun H, Hauke P, Yang B, Yuan Z~S, Berges
  J and Pan J~W 2022 {\em Science\/} {\bf 377} 311--314 (\textit{Preprint}
  \eprint{https://www.science.org/doi/pdf/10.1126/science.abl6277})
  \urlprefix\url{https://www.science.org/doi/abs/10.1126/science.abl6277}

\bibitem{halimeh2023cold}
Halimeh J~C, Aidelsburger M, Grusdt F, Hauke P and Yang B 2023
  (\textit{Preprint} \eprint{2310.12201})

\bibitem{surace2023ab}
Surace F~M, Fromholz P, Oppong N~D, Dalmonte M and Aidelsburger M 2023 {\em PRX
  Quantum\/} {\bf 4} 020330
  \urlprefix\url{https://doi.org/10.1103/PRXQuantum.4.020330}

\bibitem{surace2023scalable}
Surace F~M, Fromholz P, Scazza F and Dalmonte M 2023 Scalable, ab initio
  protocol for quantum simulating su($n$)$\times$u(1) lattice gauge theories
  (\textit{Preprint} \eprint{2310.08643})

\bibitem{schweizer2019floquet}
Schweizer C, Grusdt F, Berngruber M, Barbiero L, Demler E, Goldman N, Bloch I
  and Aidelsburger M 2019 {\em Nature Physics\/} {\bf 15} 1168--1173
  \urlprefix\url{https://doi.org/10.1038/s41567-019-0649-7}

\bibitem{weimer2010rydberg}
Weimer H, M{\"u}ller M, Lesanovsky I, Zoller P and B{\"u}chler H~P 2010 {\em
  Nature Physics\/} {\bf 6} 382--388
  \urlprefix\url{https://doi.org/10.1038/nphys1614}

\bibitem{celi2020emerging}
Celi A, Vermersch B, Viyuela O, Pichler H, Lukin M~D and Zoller P 2020 {\em
  Physical Review X\/} {\bf 10} 021057
  \urlprefix\url{https://doi.org/10.1103/PhysRevX.10.021057}

\bibitem{surace2020lattice}
Surace F~M, Mazza P~P, Giudici G, Lerose A, Gambassi A and Dalmonte M 2020 {\em
  Physical Review X\/} {\bf 10} 021041
  \urlprefix\url{https://doi.org/10.1103/PhysRevX.10.021041}

\bibitem{davoudi2020towards}
Davoudi Z, Hafezi M, Monroe C, Pagano G, Seif A and Shaw A 2020 {\em Physical
  Review Research\/} {\bf 2} 023015
  \urlprefix\url{https://doi.org/10.1103/PhysRevResearch.2.023015}

\bibitem{buazuavan2023synthetic}
B{\u{a}}z{\u{a}}van O, Saner S, Tirrito E, Araneda G, Srinivas R and Bermudez A
  2023  (\textit{Preprint} \eprint{2305.08700})

\bibitem{nguyen2022digital}
Nguyen N~H, Tran M~C, Zhu Y, Green A~M, Alderete C~H, Davoudi Z and Linke N~M
  2022 {\em PRX Quantum\/} {\bf 3}(2) 020324
  \urlprefix\url{https://link.aps.org/doi/10.1103/PRXQuantum.3.020324}

\bibitem{atas2021su2}
Atas Y~Y, Zhang J, Lewis R, Jahanpour A, Haase J~F and Muschik C~A 2021 {\em
  Nature communications\/} {\bf 12} 6499
  \urlprefix\url{https://doi.org/10.1038/s41467-021-26825-4}

\bibitem{atas2023simulating}
Atas Y~Y, Haase J~F, Zhang J, Wei V, Pfaendler S~M~L, Lewis R and Muschik C~A
  2023 {\em Phys. Rev. Res.\/} {\bf 5}(3) 033184
  \urlprefix\url{https://link.aps.org/doi/10.1103/PhysRevResearch.5.033184}

\bibitem{wang2022observation}
Wang Z, Ge Z~Y, Xiang Z, Song X, Huang R~Z, Song P, Guo X~Y, Su L, Xu K, Zheng
  D and Fan H 2022 {\em Phys. Rev. Res.\/} {\bf 4}(2) L022060
  \urlprefix\url{https://link.aps.org/doi/10.1103/PhysRevResearch.4.L022060}

\bibitem{mildenberger2022probing}
Mildenberger J, Mruczkiewicz W, Halimeh J~C, Jiang Z and Hauke P 2022
  (\textit{Preprint} \eprint{2203.08905})

\bibitem{surace2021scattering}
Surace F~M and Lerose A 2021 {\em New Journal of Physics\/} {\bf 23} 062001
  \urlprefix\url{https://doi.org/10.1088/1367-2630/abfc40}

\bibitem{domanti2023coherence}
Domanti E~C, Castorina P, Zappalà D and Amico L 2023 Coherence of confined
  matter in lattice gauge theories at the mesoscopic scale (\textit{Preprint}
  \eprint{2304.12713})

\bibitem{kumar2021cavity}
Kumar P, Biswas T, Feliz K, Kanamoto R, Chang M~S, Jha A~K and Bhattacharya M
  2021 {\em Phys. Rev. Lett.\/} {\bf 127}(11) 113601
  \urlprefix\url{https://link.aps.org/doi/10.1103/PhysRevLett.127.113601}

\bibitem{pradhan2023cavity}
{Pradhan} N, {Kumar} P, {Kanamoto} R, {Nath Dey} T, {Bhattacharya} M and
  {Mishra} P~K 2023 {\em arXiv e-prints\/} arXiv:2306.06720 (\textit{Preprint}
  \eprint{2306.06720})

\bibitem{zhang2019magnetic}
Zhang B, Siercke M, Chan K~S, Beian M, Lim M~J and Dumke R 2012 {\em Phys. Rev.
  A\/} {\bf 85}(1) 013404
  \urlprefix\url{https://link.aps.org/doi/10.1103/PhysRevA.85.013404}

\bibitem{muller2010trapping}
Müller T, Zhang B, Fermani R, Chan K~S, Wang Z~W, Zhang C~B, Lim M~J and Dumke
  R 2010 {\em New Journal of Physics\/} {\bf 12} 043016
  \urlprefix\url{https://dx.doi.org/10.1088/1367-2630/12/4/043016}

\bibitem{cano2009meissner}
Cano D, Kasch B, Hattermann H, Kleiner R, Zimmermann C, Koelle D and Fort\'agh
  J 2008 {\em Phys. Rev. Lett.\/} {\bf 101}(18) 183006
  \urlprefix\url{https://link.aps.org/doi/10.1103/PhysRevLett.101.183006}

\bibitem{mukai2007persistent}
Mukai T, Hufnagel C, Kasper A, Meno T, Tsukada A, Semba K and Shimizu F 2007
  {\em Phys. Rev. Lett.\/} {\bf 98}(26) 260407
  \urlprefix\url{https://link.aps.org/doi/10.1103/PhysRevLett.98.260407}

\bibitem{chormaic2023probing}
Chormaic S~N, Vylegzhanin A, Shahrabifarahani Z, Raj A, Zaitsev A, Abdrakhmanov
  S, Li W, Everett J~L and Brown D 2023 Probing cold atom interactions via
  optical nanofibers {\em Quantum Sensing, Imaging, and Precision Metrology\/}
  ed Shahriar S~M and Scheuer J ({SPIE})
  \urlprefix\url{https://doi.org/10.1117/12.2657345}

\bibitem{russell2013laser}
Russell L, Kumar R, Tiwari V and {Nic Chormaic} S 2013 {\em Optics
  Communications\/} {\bf 309} 313--317 ISSN 0030-4018
  \urlprefix\url{https://www.sciencedirect.com/science/article/pii/S0030401813007189}

\bibitem{goban2012demonstration}
Goban A, Choi K~S, Alton D~J, Ding D, Lacro\^ute C, Pototschnig M, Thiele T,
  Stern N~P and Kimble H~J 2012 {\em Phys. Rev. Lett.\/} {\bf 109}(3) 033603
  \urlprefix\url{https://link.aps.org/doi/10.1103/PhysRevLett.109.033603}

\bibitem{sague2007cold}
Sagu\'e G, Vetsch E, Alt W, Meschede D and Rauschenbeutel A 2007 {\em Phys.
  Rev. Lett.\/} {\bf 99}(16) 163602
  \urlprefix\url{https://link.aps.org/doi/10.1103/PhysRevLett.99.163602}

\bibitem{westbrook1998new}
Westbrook N, Westbrook C~I, Landragin A, Labeyrie G, Cognet L, Savalli V,
  Horvath G, Aspect A, Hendel C, Moelmer K, Courtois J~Y, Phillips W~D, Kaiser
  R and Bagnato V 1998 {\em Physica Scripta\/} {\bf 1998} 7
  \urlprefix\url{https://dx.doi.org/10.1238/Physica.Topical.078a00007}

\bibitem{gillen2009twodimensional}
Gillen J~I, Bakr W~S, Peng A, Unterwaditzer P, F\"olling S and Greiner M 2009
  {\em Phys. Rev. A\/} {\bf 80}(2) 021602
  \urlprefix\url{https://link.aps.org/doi/10.1103/PhysRevA.80.021602}

\bibitem{schloss2020controlled}
Schloss J, Barnett P, Sachdeva R and Busch T 2020 {\em Phys. Rev. A\/} {\bf
  102}(4) 043325
  \urlprefix\url{https://link.aps.org/doi/10.1103/PhysRevA.102.043325}

\bibitem{hejazi2020symmetry}
Hejazi S~S~S, Polo J, Sachdeva R and Busch T 2020 {\em Phys. Rev. A\/} {\bf
  102}(5) 053309
  \urlprefix\url{https://link.aps.org/doi/10.1103/PhysRevA.102.053309}

\bibitem{hejazi2022formation}
Hejazi S~S~S, Polo J and Tsubota M 2022 {\em Phys. Rev. A\/} {\bf 105}(5)
  053307 \urlprefix\url{https://link.aps.org/doi/10.1103/PhysRevA.105.053307}

\bibitem{fleischhauer2005electromagnetically}
Fleischhauer M, Imamoglu A and Marangos J~P 2005 {\em Reviews of Modern
  Physics\/} {\bf 77} 633–673 ISSN 1539-0756
  \urlprefix\url{http://dx.doi.org/10.1103/RevModPhys.77.633}

\bibitem{hammerer2010quantum}
Hammerer K, Sørensen A~S and Polzik E~S 2010 {\em Reviews of Modern Physics\/}
  {\bf 82} 1041–1093 ISSN 1539-0756
  \urlprefix\url{http://dx.doi.org/10.1103/RevModPhys.82.1041}

\bibitem{xiang_hybrid_2013}
Xiang Z~L, Ashhab S, You J~Q and Nori F 2013 {\em Reviews of Modern Physics\/}
  {\bf 85} 623–653 ISSN 1539-0756
  \urlprefix\url{http://dx.doi.org/10.1103/RevModPhys.85.623}

\bibitem{yu2016charge}
Yu D, Valado M~M, Hufnagel C, Kwek L~C, Amico L and Dumke R 2016 {\em Physical
  Review A\/} {\bf 93} 042329
  \urlprefix\url{https://doi.org/10.1103/PhysRevA.93.042329}

\bibitem{yu2017superconducting}
Yu D, Kwek L~C, Amico L and Dumke R 2017 {\em Quantum Science and Technology\/}
  {\bf 2} 035005 \urlprefix\url{https://doi.org/10.1088/2058-9565/aa7c50}

\bibitem{yu2016superconducting}
Yu D, Landra A, Valado M~M, Hufnagel C, Kwek L~C, Amico L and Dumke R 2016 {\em
  Physical Review A\/} {\bf 94} 062301
  \urlprefix\url{https://doi.org/10.1103/PhysRevA.94.062301}

\bibitem{yu2016quantum}
Yu D, Valado M~M, Hufnagel C, Kwek L~C, Amico L and Dumke R 2016 {\em
  Scientific Reports\/} {\bf 6} 38356
  \urlprefix\url{https://doi.org/10.1038/srep38356}

\bibitem{yu2018stabilizing}
Yu D, Landra A, Kwek L~C, Amico L and Dumke R 2018 {\em New Journal of
  Physics\/} {\bf 20} 023031
  \urlprefix\url{https://doi.org/10.1088/1367-2630/aaa643}

\bibitem{yu2017theoretical}
Yu D, Kwek L~C, Amico L and Dumke R 2017 {\em Physical Review A\/} {\bf 95}
  053811 \urlprefix\url{https://doi.org/10.1103/PhysRevA.95.053811}

\bibitem{yu2018charge}
Yu D, Kwek L~C, Amico L and Dumke R 2018 {\em Physical Review A\/} {\bf 98}(3)
  033833 \urlprefix\url{https://doi.org/10.1103/PhysRevA.98.033833}

\bibitem{cano2008meissner}
Cano D, Kasch B, Hattermann H, Kleiner R, Zimmermann C, Koelle D and Fortágh J
  2008 {\em Physical Review Letters\/} {\bf 101} ISSN 1079-7114
  \urlprefix\url{http://dx.doi.org/10.1103/PhysRevLett.101.183006}

\bibitem{nirrengarten2006realization}
Nirrengarten T, Qarry A, Roux C, Emmert A, Nogues G, Brune M, Raimond J~M and
  Haroche S 2006 {\em Physical Review Letters\/} {\bf 97} 200405
  \urlprefix\url{https://doi.org/10.1103/PhysRevLett.97.200405}

\bibitem{tosto2019optically}
Tosto F, Baw~Swe P, Nguyen N~T, Hufnagel C, Mart\'{\i}nez~Valado M, Prigozhin
  L, Sokolovsky V and Dumke R 2019 {\em Applied Physics Letters\/} {\bf 114}
  222601 \urlprefix\url{https://doi.org/10.1063/1.5096997}

\bibitem{muller2010programmable}
M\"{u}ller T, Zhang B, Fermani R, Chan K, Lim M and Dumke R 2010 {\em Physical
  Review A\/} {\bf 81} 053624
  \urlprefix\url{https://doi.org/10.1103/PhysRevA.81.053624}

\bibitem{PhysRevA.85.013404}
Zhang B, Siercke M, Chan K~S, Beian M, Lim M~J and Dumke R 2012 {\em Phys. Rev.
  A\/} {\bf 85} 013404

\bibitem{verdu2009strong}
Verd\'{u} J, Zoubi H, Koller C, Majer J, Ritsch H and Schmiedmayer J 2009 {\em
  Physical Review Letters\/} {\bf 103} 043603
  \urlprefix\url{https://doi.org/10.1103/PhysRevLett.103.043603}

\bibitem{petrosyan2019microwave}
Petrosyan D, M\o{}lmer K, Fort\'{a}gh J and Saffman M 2019 {\em New Journal of
  Physics\/} {\bf 21} 073033
  \urlprefix\url{https://doi.org/10.1088/1367-2630/ab307c}

\bibitem{bernon2013manipulation}
Bernon S, Hattermann H, Bothner D, Knufinke M, Weiss P, Jessen F, Cano D,
  Kemmler M, Kleiner R, Koelle D {\em et~al.\/} 2013 {\em Nature
  Communications\/} {\bf 4} 1--8
  \urlprefix\url{https://doi.org/10.1038/ncomms3380}

\bibitem{hattermann2017coupling}
Hattermann H, Bothner D, Ley L, Ferdinand B, Wiedmaier D, S\'{a}rk\'{a}ny L,
  Kleiner R, Koelle D and Fort\'{a}gh J 2017 {\em Nature communications\/} {\bf
  8} 1--7 \urlprefix\url{https://doi.org/10.1038/s41467-017-02439-7}

\bibitem{fortagh2007magnetic}
Fort\'{a}gh J and Zimmermann C 2007 {\em Reviews of Modern Physics\/} {\bf 79}
  235--289 \urlprefix\url{https://doi.org/10.1103/RevModPhys.79.235}

\bibitem{bao2012efficient}
Bao X~H, Reingruber A, Dietrich P, Rui J, D\"{u}ck A, Strassel T, Li L, Liu
  N~L, Zhao B and Pan J~W 2012 {\em Nature Physics\/} {\bf 8} 517--521
  \urlprefix\url{https://doi.org/10.1038/nphys2324}

\bibitem{cho2016coherent}
Cho Y~W, Campbell G~T, Everett J~L, Bernu J, Higginbottom D~B, Cao M~T, Geng J,
  Robins N~P, Lam P~K and Buchler B~C 2016 {\em Optica\/} {\bf 3} 100--107
  \urlprefix\url{https://opg.optica.org/optica/abstract.cfm?URI=optica-3-1-100}

\bibitem{wigley2016fast}
Wigley P~B, Everitt P~J, van~den Hengel A, Bastian J~W, Sooriyabandara M~A,
  McDonald G~D, Hardman K~S, Quinlivan C~D, Manju P, Kuhn C~C~N, Petersen I~R,
  Luiten A~N, Hope J~J, Robins N~P and Hush M~R 2016 {\em Scientific Reports\/}
  {\bf 6} \urlprefix\url{https://doi.org/10.1038/srep25890}

\bibitem{barker2020applying}
Barker A~J, Style H, Luksch K, Sunami S, Garrick D, Hill F, Foot C~J and
  Bentine E 2020 {\em Machine Learning: Science and Technology\/} {\bf 1}
  015007 \urlprefix\url{https://dx.doi.org/10.1088/2632-2153/ab6432}

\bibitem{ness2020single}
Ness G, Vainbaum A, Shkedrov C, Florshaim Y and Sagi Y 2020 {\em Phys. Rev.
  Appl.\/} {\bf 14}(1) 014011
  \urlprefix\url{https://link.aps.org/doi/10.1103/PhysRevApplied.14.014011}

\bibitem{metz2021deep}
Metz F, Polo J, Weber N and Busch T 2021 {\em Machine Learning: Science and
  Technology\/} {\bf 2} 035019
  \urlprefix\url{https://dx.doi.org/10.1088/2632-2153/abea6a}

\bibitem{kim2023vortex}
Kim M, Kwon J, Rabga T and Shin Y 2023 {\em Machine Learning: Science and
  Technology\/} {\bf 4} 045017
  \urlprefix\url{https://dx.doi.org/10.1088/2632-2153/ad03ad}

\bibitem{chih2021reinforcement}
Chih L~Y and Holland M 2021 {\em Phys. Rev. Res.\/} {\bf 3}(3) 033279
  \urlprefix\url{https://link.aps.org/doi/10.1103/PhysRevResearch.3.033279}

\bibitem{abad2014persistent}
Abad M, Sartori A, Finazzi S and Recati A 2014 {\em Physical Review A\/} {\bf
  89}(5) 053602

\bibitem{simjanovski2023optimizing}
{Simjanovski} S, {Gauthier} G, {Davis} M~J, {Rubinsztein-Dunlop} H and {Neely}
  T~W 2023 {\em arXiv e-prints\/} arXiv:2304.06199 (\textit{Preprint}
  \eprint{2304.06199})

\bibitem{north2023albert}
North H, Fitch N~J and Tingle A~E 2023 Albert: a cloud-based quantum-sensor
  development platform for the masses {\em Quantum Sensing, Imaging, and
  Precision Metrology\/} ed Shahriar S~M and Scheuer J (SPIE)
  \urlprefix\url{http://dx.doi.org/10.1117/12.2650576}

\bibitem{huber2022cloud}
Huber F, Amato-Grill J, Bylinskii A, Cantu S~H, Hu M~G, Kim D, Lukin A, Gemelke
  N and Keesling A 2022 Cloud-accessible, programmable quantum simulator based
  on two-dimensional neutral atom arrays {\em Quantum 2.0 Conference and
  Exhibition\/} (Optica Publishing Group) p QW3A.2
  \urlprefix\url{https://opg.optica.org/abstract.cfm?URI=QUANTUM-2022-QW3A.2}

\end{thebibliography}

\end{document}